# A Supercooled Spin Liquid State
# in the Frustrated Pyrochlore Dy$_2$Ti$_2$O$_7$

Ethan R. Kassner[1], Azar B. Eyvazov[1], Benjamin Pichler[1,2], Timothy J. S. Munsie[3,4],
Hanna A. Dabkowska[3], Graeme M. Luke[3,4,5] & J.C. Séamus Davis[1,6,7,8]

[1] *LASSP, Department of Physics, Cornell University, Ithaca, NY 14853, USA*
[2] *Department of Physics, Stanford University, Palo Alto, CA 94301, USA*
[3] *Brockhouse Institute for Materials Research, McMaster University, Hamilton, ON, Canada.*
[4] *Department of Physics, McMaster University, Hamilton, Ontario, L8S 4M1, Canada.*
[5] *Canadian Institute for Advanced Research, Toronto, Ontario, M5G 1Z8, Canada*
[6] *CMPMS Department, Brookhaven National Laboratory, Upton, NY 11973, USA.*
[7] *School of Physics , University of St. Andrews, St. Andrews, Fife KY16 9SS, Scotland.*
[8] *Kavli Institute at Cornell for Nanoscale Science, Cornell University, Ithaca, NY 14850, USA.*

**A 'supercooled' liquid develops when a fluid does not crystallize upon cooling below its ordering temperature. Instead, the microscopic relaxation times diverge so rapidly that, upon further cooling, equilibration eventually becomes impossible and glass formation occurs. Classic supercooled liquids exhibit specific identifiers including microscopic relaxation times diverging on a Vogel-Tammann-Fulcher (VTF) trajectory, a Havriliak-Negami (HN) form for the dielectric function $\varepsilon(\omega, T)$, and a general Kohlrausch-Williams-Watts (KWW) form for time-domain relaxation. Recently, the pyrochlore Dy$_2$Ti$_2$O$_7$ has become of interest because its frustrated magnetic interactions may, in theory, lead to highly exotic magnetic fluids. However, its true magnetic state at low temperatures has proven very difficult to identify unambiguously. Here we introduce high-precision, boundary-free magnetization transport techniques based upon toroidal geometries and gain a fundamentally new understanding of the time- and frequency-dependent magnetization dynamics of Dy$_2$Ti$_2$O$_7$. We demonstrate a virtually universal HN form for the magnetic susceptibility $\chi(\omega, T)$, a general KWW form for the real-time magnetic relaxation, and a divergence of the microscopic magnetic relaxation rates with precisely the VTF trajectory. Low temperature Dy$_2$Ti$_2$O$_7$ therefore exhibits the characteristics of a supercooled magnetic liquid; the consequent implication is that this translationally invariant lattice of strongly correlated spins is evolving towards an unprecedented magnetic glass state, perhaps due to many-body localization of spin.**

## Supercooled Liquids

Cooling a pure liquid usually results in crystallization via a first order phase transition. However, in glass-forming liquids when the cooling is sufficiently rapid, a metastable 'supercooled' state is achieved instead (*1-3*). Here, the microscopic relaxation times diverge until equilibration of the system is no longer possible at a given cooling rate. At this juncture there is generally a broad peak in the specific heat preceding the glass transition, at which no symmetry breaking phase transition occurs (Fig. 1A). The antecedent fluid exhibits a set of phenomena characteristic of the supercooled liquid state (1-*3*). For example, the divergence of microscopic relaxation times $\tau_0(T)$ typically shows substantial departures from Arrhenius behavior ($\tau_0(T) = A\exp(\Delta/kT)$) and, instead, is described characteristically using the Vogel-Tammann-Fulcher (VTF) form (*4*)

$$\tau_0(T) = A\exp\left(\frac{DT_0}{T-T_0}\right) \qquad (1)$$

Here $T_0$ is a temperature at which the relaxation time diverges to $\infty$ while $D$ characterizes the extent of the super-Arrhenius behavior (Fig. 1B). One way to establish $\tau_0(T)$ is by measuring the characteristic frequency $\omega_0(T) = 1/\tau_0(T)$ of peaks in the dissipative (imaginary) component of the dielectric function $\varepsilon(\omega, T)$. For classic supercooled liquids, $\varepsilon(\omega, T)$ generally exhibits the Havriliak-Negami (HN ) form (*5,6*)

$$\varepsilon(\omega, T) = \varepsilon_\infty + \frac{\varepsilon_0}{(1+(i\omega\tau_{HN})^\alpha)^\gamma} \qquad (2)$$

Here the exponents α and γ describe, respectively, the broadening and asymmetry of the relaxation in frequency compared to a simple Debye form (α=γ=1), $\tau_{HN}(T)$ is the microscopic relaxation time, and $\epsilon_\infty$ is a purely real quantity that describes the relaxation in the $\omega \to \infty$ limit. The HN forms of $Re[\varepsilon(\omega, T)]$ and $Im[\varepsilon(\omega, T)]$ that are characteristic of supercooled liquids are shown in Fig. 1C. In the time domain, this relaxation is described by the classic Kohlrausch-Williams-Watts (KWW) form (*7*)

$$\varepsilon(t) = \varepsilon_0\exp\left[-\left(\frac{t}{\tau_{KWW}}\right)^\beta\right] \qquad (3)$$

where $\varepsilon(t)$ describes the evolution of polarization **P**(t), $\tau_{KWW}$ is the microscopic relaxation time, and 0< $\beta$ <1 is a "stretching exponent" (Fig. 1D). Debye relaxation corresponds to $\beta = 1$ while $\beta < 1$ typically indicates the presence of a more complex energy barrier distribution. Although the KWW function has no analytic Fourier transform when $\beta \neq 1/2$, KWW and HN are actually complementary descriptions of the same microscopic phenomena, being connected by the relations (*8*)

$$\ln\left(\frac{\tau_{HN}}{\tau_{KWW}}\right) = 2.6(1-\beta)^{0.5}\exp(-3\beta) \text{ with } \alpha\gamma = \beta^{1.23} \quad (4)$$

Observation of this HN / KWW phenomenology is the generally used standard by which classic supercooled fluids are identified (*1,2,3,4*).

**Magnetization Dynamics Studies of $Dy_2Ti_2O_7$**

Frustrated magnetic pyrochlores are now the focus of widespread interest because of the possibility that they can support different exotic magnetic phases (*9-14*). The pyrochlore $Dy_2Ti_2O_7$ is one of the most widely studied; it consists of highly-magnetic $Dy^{3+}$ ions in a sub-lattice comprised of corner-sharing tetrahedra (Fig. 2A), and an interpenetrating octahedral lattice of $Ti^{4+}$ ions playing no magnetic role. Crystal fields break the angular-momentum-state degeneracy and cause the $Dy^{3+}$ moments ($J = \frac{15}{2}, \mu \approx 10\mu_B$) to point along their local [111] directions (*15*). Although high-temperature susceptibility measurements indicate a Curie-Weiss temperature of ~1.2K (*16*), both low temperature susceptibility (*16*) and muon spin rotation studies (*17*) have revealed no magnetic ordering transition in $Dy_2Ti_2O_7$ down to T~50 mK. Additionally, a broad peak in the specific heat (*18-22*) centered around $T \approx 1.0$ K occurs at or below the transition temperature that might be expected from the Curie-Weiss temperature, but no phase transition occurs. By contrast, in a typical paramagnet the heat capacity would exhibit a sharp peak at the ordering temperature below which long-range magnetic order would become apparent in for example the susceptibility, muon spin rotation, and neutron scattering; none of these ordering indications are observed in $Dy_2Ti_2O_7$. Instead, dipole and exchange interactions combine to create an effective nearest-neighbor Ising interaction of the form $-J_{eff}\sum \boldsymbol{S}_i \cdot \boldsymbol{S}_j$ with $J_{eff} \approx 1.1$ K (*23*). A consequence of this interaction in the pyrochlore geometry is that there are six possible equivalent magnetic ground state conformations of a single $Dy_2Ti_2O_7$ tetrahedron; these can be mapped to the Bernal-Fowler ("2-in, 2-out") rules that govern hydrogen bond configurations in water ice but now it is a 2-in, 2-out arrangement of $Dy^{3+}$ moments (Fig. 2B). This elegant 'spin-ice' configuration has been firmly established (*10,14,21,22,24*). Theoretically, the long range dipole interactions can also generate magnetic ordering (*10,14,25*) but, significantly, this has not been observed down to temperatures below 50mK.

The magnetic excited states of $Dy_2Ti_2O_7$ are then of great interest. One conjecture is that the magnetization dynamics may be described as a fluid of emergent delocalized magnetic monopoles (*26*). The widely-used dipolar spin-ice model (DSIM) (*23*) can be used to derive this magnetic monopoles in spin-ice (MMSI) picture. DSIM incorporates nearest-neighbor exchange interactions and long-range dipole interactions:

$$H_{DSI} = -J \sum_{<ij>} \boldsymbol{S}_i \cdot \boldsymbol{S}_j + Da^3 \sum_{i<j} \left( \frac{\boldsymbol{S}_i \cdot \boldsymbol{S}_j}{|\boldsymbol{r}_{ij}|^3} - \frac{3(\boldsymbol{S}_i \cdot \boldsymbol{r}_{ij})(\boldsymbol{S}_j \cdot \boldsymbol{r}_{ij})}{|\boldsymbol{r}_{ij}|^5} \right) \tag{5}$$

where $a$ is the nearest-neighbor distance between moments, $J$ is an exchange strength, and $D = \mu_0 \mu^2/(4\pi a^3)$ is a nearest-neighbor dipole energy scale. This Hamiltonian is mathematically equivalent to a model where flips of the real $Dy^{3+}$ dipoles are recast as two opposite magnetic charges which, though a sequence of spin flips, are hypothesized to form a fluid of delocalized magnetic monopoles (red and green in Fig. 2C) (*26*). At low temperatures these monopoles might then form a dilute neutral gas analogous to an electrolyte, so that a Debye-Hückel electrolyte model (*27*) suitably modified for magnetic monopoles, may be used to describe a 'magnetolyte' state (*28*). Such a mobile fluid of magnetic monopole excitations, if extant, would constitute a highly novel magnetic state.

However, many observed properties of $Dy_2Ti_2O_7$ remain unexplained when they are analyzed using the DSIM/MMSI models. Though DSIM captures some of the diffuse neutron scattering features of $Dy_2Ti_2O_7$, simulations based on Eqn. (5) give an incomplete description of the data; additional exchange contributions from next-nearest and third-nearest neighbors are required to actually fit the measured scattering intensities precisely (*29*). Similarly, while the magnetic susceptibility $\chi(\omega, T) = \chi'(\omega, T) - i\chi''(\omega, T)$ is known empirically with high precision (*30-33*) no accurate quantitative model based on internally consistent parameterizations of $\chi'(\omega, T)$ and $\chi''(\omega, T)$ has been achieved using DSIM/MMSI (or any other) models. Moreover, the measured susceptibility of $Dy_2Ti_2O_7$ cannot be described by the simple Debye form (e.g Eqn. 2 with α=γ=1) that would be expected of a typical paramagnet (*33*). Microscopic approaches have also failed to capture precisely the $Dy_2Ti_2O_7$ magnetization dynamics. For example, DSIM simulations do not accurately reproduce the measured $\chi(\omega, T)$ data at low temperatures (*34*). Furthermore, below $\sim 0.5K$ macroscopic relaxation rates in $Dy_2Ti_2O_7$ become ultra-slow (*20,21,30,33,35,36,37*); this reflects a divergence of microscopic relaxation times (*30,33*) that is unexplained quantitatively within the present DSIM/MMSI models. Thus, although Debye-Hückel calculations (*28*) and DSIM/MMSI simulations (*38,39*) offer clear improvements over a simple Arrhenius form for $\tau(T)$, they still significantly underestimate the observed magnitude and rate of increase of $\tau$ below 1 K. In fact, as emphasized in a recent review (*14*), no studies of $Dy_2Ti_2O_7$ have yielded direct and unambiguous evidence of a fluid of delocalized magnetic monopoles. This motivates the search for a more accurate identification and understanding of the low temperature magnetic state of this important compound.

**Experimental Methods**

To explore these issues, we introduce a novel high-precision, boundary-free technique for studying magnetization transport in $Dy_2Ti_2O_7$. The innovation consists primarily of using a toroidal geometry for both the $Dy_2Ti_2O_7$ samples and the magnetization sensors, an arrangement with several important benefits. The first is that the superconducting toroidal solenoid (STS) can be used to both drive magnetization flows azimuthally and to simultaneously and directly detect d***M***/dt throughout the whole torus. More significantly, this topology removes any boundaries in the direction of the magnetization transport. To achieve this sample topology, holes were pierced through disks of single-crystal $Dy_2Ti_2O_7$ (SI Section 1). A superconducting toroidal solenoid of NbTi wire (diameter$\sim$ 0.1 mm) is then wound around each $Dy_2Ti_2O_7$ torus (SI Section 1). Typical samples had an ID $\sim$ 2.5 mm, OD $\sim$ 6 mm, and thickness $\sim$ 1 mm. The superconducting circuitry used to drive and measure d***M***/dt is shown schematically in Fig. 2D, along with indicators of the azimuthal $B_\phi$ generated by the STS (blue arrows) and the putative flow of a mixed-sign monopole fluid (red/green arrows). The all-superconducting 4-point I-V measurement circuit used to generate the $B\phi$-field and simultaneously measure the EMF induced by d$\boldsymbol{M}$/dt is shown schematically in Fig. 2D. These toroidal sample/coil assemblies were mounted on a dilution refrigerator and studied at temperatures 30mK<T<3K, and using currents not exceeding 30 mA in the STS ($B\phi < \sim 1$ G, so always in the low-field limit $\mu B \ll k_B T$).

**Time-domain Measurements**

Elementary magnetization dynamics experiments in the time-domain are then carried out using the following repeated measurement cycle. First, a current (and thus $B\phi$) is switched on in the STS and the EMF across this same coil $V$(t) is measured from when the switch-on transient ends until $V$(t) drops below our noise level. Next, the current (and $B\phi$) is turned off and we again measure the $V$(t) response. The current is then turned on in the reverse direction (-$B\phi$) for the same time interval and turned off for the same time interval, and the procedure is repeated. Complete data sets for this sequence as a function of temperature are shown in SI Section 3.

Fig. 3A shows the time-dependent magnetization of Dy$_2$Ti$_2$O$_7$ determined from $V(t,T) \propto d\boldsymbol{M}/dt[t,T]$ at each temperature (Fig. 3A). To study these measurements in the context of KWW, we fit each transient $V(t,T)$ to the KWW function Eqn. 3 and determine $\tau_{KWW}(T)$ and $\beta(T)$. The results are shown directly by plotting the measured $\log\frac{V(t)}{V_0}$ versus $x(t) \equiv ((t)/\tau_{KWW}(T))^{\beta(T)}$ in Fig. 3B; for these temperatures $\beta \sim 0.75$. There we see that, for all measured transients over a range 575mK < T < 900mK, the $d\boldsymbol{M}/dt$ of Dy$_2$Ti$_2$O$_7$ is very well represented by a KWW time dependence. This puts the ultra-slow magnetization dynamics in the same empirical class as classic supercooled liquids (*1-4*). Stretched-exponential relaxation has been previously seen in studies of single-crystal rods of Dy$_2$Ti$_2$O$_7$ (*37*) but there it was proposed that this relaxation occurred due to the open boundary conditions. However, our sample topology is a physical realization of periodic boundary conditions implying that our observed KWW relaxation is actually a fundamental property of the system. Thus, while simulations of the DSI and delocalized MMSI models in periodic boundary conditions explicitly rule out stretched-exponential relaxation of magnetization in Dy$_2$Ti$_2$O$_7$ (39), our measurement results exhibit these characteristics comprehensively.

**Frequency-domain measurements**

If KWW magnetization transient dynamics as in Fig. 3 are evidence for supercooling of a correlated magnetic fluid, the equivalent phenomena should also be observed as a universal HN form for the frequency-dependent complex magnetic susceptibility $\chi(\omega, T)$. To explore this possibility, we apply sinusoidal currents $I_0 \text{Cos}(\omega t)$ to the STS and generate a field $B\phi \text{Cos}(\omega t + \phi)$ while simultaneously detecting the resulting EMF: $V(\omega, T) = V_x(\omega, T) + iV_y(\omega, T)$. Here the $V(\omega, T)$ is due to temperature dependent changes in the sample dynamical magnetization (SI Section 2). Directly from the experimental setup (SI Section 2) one can write $V_x(\omega, T) = -I_0 \omega L \chi''$ , $V_y(\omega, T) = -I_0 \omega L \chi'$ where $L$ is the effective geometrical inductance of the STS. A complete set of $\chi'(\omega, T)$ and $\chi''(\omega, T)$ measured in this fashion is shown in Fig 4A,B and described in SI Section 2. Equivalent phenomena were measured in three different $Dy_2Ti_2O_7$ tori ruling out any specific sample preparation or torus/coil geometry effects as the cause of reported phenomena. Empirically, the temperature and frequency dependence of the $\chi'(\omega, T)$ and $\chi''(\omega, T)$ in our studies is virtually identical to what has been reported previously. Until now, however, no quantitative internally consistent model for the susceptibility of $Dy_2Ti_2O_7$ has been identified. Here, we show that a HN form for a magnetic susceptibility

$$\chi_{HN}(\omega, T) = \chi_\infty + \frac{\chi_0}{(1 + (i\omega\tau_{HN})^\alpha)^\gamma} \qquad (6)$$

provides a comprehensive accurate internally consistent description for both $\chi'(\omega, T)$ and $\chi''(\omega, T)$ of $Dy_2Ti_2O_7$. To achieve this, we need to demonstrate that all our disparate $\chi(\omega, T)$ data has the same HN functional form. We define a scaled susceptibility $G(\gamma, \chi)$ that depends only upon $\gamma$, $\chi_0$ and $\chi(\omega, T)$:

$$\text{Re}[G(\gamma, \chi)] \equiv \chi_0^{-1/\gamma}[(\chi')^2 + (\chi'')^2]^{\frac{1}{2\gamma}} \cos\left[\frac{1}{\gamma}\arctan\left(\frac{\chi''}{\chi'}\right)\right] \qquad (7)$$

$$\text{Im}[G(\gamma, \chi)] \equiv \chi_0^{-1/\gamma}[(\chi')^2 + (\chi'')^2]^{\frac{1}{2\gamma}} \sin\left[\frac{1}{\gamma}\arctan\left(\frac{\chi''}{\chi'}\right)\right]$$

where we have neglected $\chi_\infty$ because it is much smaller than $\chi_0$ at our experimental temperatures (SI Section 2). As shown analytically in SI Section 2, if plotting $\text{Re}[G(\gamma, \chi)]$ and $\text{Im}[G(\gamma, \chi)]$ vs. $(\omega\tau)^\alpha$ collapses $\chi(\omega, T)$ data onto a single curve, then the magnetic susceptibility exhibits a universal HN form. Figure 4C,D show this data collapse for all measured $\chi(\omega, T)$ where the fine solid line is the model HN form of Eqn. 6. It is clear that $\chi_{HN}(\omega, T)$ fits the observed $Dy_2Ti_2O_7$ data precisely and comprehensively. To our knowledge, this is the first time that both $\chi'(\omega, T)$ and $\chi''(\omega, T)$ of $Dy_2Ti_2O_7$ have been quantitatively and simultaneously described by a single internally-consistent function, across very wide frequency/temperature ranges and with Kramers-Kronig consistency. More importantly, since this HN form for magnetic susceptibility is functionally indistinguishable from the $\varepsilon(\omega, T)$ of supercooled liquids (*1-3*) it strongly implies supercooling of a correlated magnetic fluid in this compound.

**Diverging Microscopic Relaxation Rates**

A final test of the supercooled magnetic liquid hypothesis would be to show that the magnetic relaxation times diverge on a VTF trajectory for $Dy_2Ti_2O_7$. To determine microscopic relaxation times $\tau_0(T)$ spanning the whole temperature range, we use $\tau_{HN}(T)$ from fitting Eqn. 6 simultaneously to $\chi'(\omega, T)$ and $\chi''(\omega, T)$ for 0.8K < T < 3K, and $\tau_{KWW}(T)$ from the time-domain V(*t,T*) fitted by Eqn. 3 and then converted to the equivalent $\tau_{HN}(T)$ for 0.58K<T<0.85K using Eqn. 4. As shown in Fig. 4E the resulting $\tau_{HN}(T)$ of $Dy_2Ti_2O_7$ can, indeed, be represented by a VTF function (Eqn. 1) with high precision over many orders of magnitude ($T_0 \approx 0.24$K and fragility parameter $D \approx 14$). This unifies the evidence (Figs. 3,4) that the magnetic state of $Dy_2Ti_2O_7$ for 0.5 K < T < 3K is a supercooled magnetic liquid.

**Conclusions and Discussion**

One may now reconsider the anomalous phenomena of $Dy_2Ti_2O_7$ in this new context. Empirically, our measured $\chi(\omega, T)$ data are in good agreement with earlier reports (*30-33*) implying that all can be describable by a HN susceptibility (Eqn. 2; Fig. 4A,B). Also, while the Curie-Weiss temperature $T_{CW} \sim 1.2$K (*16*) implies a tendency towards ferromagnetic order, no ordering is observed and, instead, a broad peak in specific heat $C(T)$ appears just below $T_{CW}$ (*18,19,20,21,22*); this is as expected for a supercooled liquid (Fig. 1A and Refs. *1-3*). The location of $Dy^{3+}$ moments in the highly anisotropic environment of the $Dy_2Ti_2O_7$ prevents ferromagnetic ordering at a temperature that might be expected from the nearest-neighbor interaction energy scale; this may be analogous to preventing the onset of a crystalline phase in glass-forming liquids due to anisotropic interactions between molecules. Moreover, ultra-slow macroscopic equilibration is widely reported at lower temperatures ([20, 35,37]) and is also is just what one expects in a supercooled liquid approaching the glass transition ([1-3]). Our observed stretched-exponential form for ultra-slow magnetization relaxation agrees well with previous studies ([37]). While actually at odds with the predictions of DSIM/MMSI simulations for periodic geometries ([39]) this phenomenon is characteristic of a supercooled fluid ([1,2,3]). Finally, the published data on divergences of microscopic relaxation times ([30,33]) are in good empirical agreement with ours, implying that the VTF form for $\tau(T)$ is general for $Dy_2Ti_2O_7$ (Eqn. 1, Fig 4E). Thus we conjecture that overall magnetization dynamics of $Dy_2Ti_2O_7$ are best explained if the system is a classical correlated-spin liquid that that is supercooled and approaching a glass transition. Within this picture, the divergence temperature $T_0 \approx 240$ mK derived from our VTF fit (Fig. 4E) provides an estimate of the lowest temperature at which a metastable magnetic liquid state can survive under arbitrary cooling protocols; below this temperature we expect that $Dy_2Ti_2O_7$ must transition into either a heterogeneous glass phase or a phase with global magnetic order.

To recapitulate: for the magnetic pyrochlore system $Dy_2Ti_2O_7$ we discover that the magnetic susceptibility $\chi(\omega, T)$ exhibits a Havriliak-Negami form, the magnetic relaxation $\chi(t)$ simultaneously exhibits the Kohlrausch-Williams-Watts form, while the microscopic magnetic relaxation rates $\tau(T)$ occur on the Vogel-Tammann-Fulcher trajectory. When, in combination with a broad specific heat peak, this phenomenology is observed for the $\varepsilon(\omega, T)$ of a classic glass forming material, it definitely identifies a supercooled liquid ([1-3]). Our observations therefore strongly imply that the magnetic state of $Dy_2Ti_2O_7$ is a supercooled classical spin liquid, approaching a glass transition. However, we emphasize that one should not expect any consequent magnetic glass to be a classic spin glass, because all the $Dy_2Ti_2O_7$ spins are at ordered crystal lattice sites with locally identical spin Hamiltonians. And, indeed, $Dy_2Ti_2O_7$ is known to exhibit a very different field-dependence from what is seen in classical spin glasses ([30]). The supercooled liquid characteristics of magnetization dynamics in $Dy_2Ti_2O_7$ (Fig.s 3,4) more likely imply some form of persistent heterogeneous freezing in the microscopic configurations of strongly correlated spins ([40,41]). Such a situation could also be described (somewhat redundantly) as freezing of monopole configurations. However, in terms of actual magnetization transport, the observed stretched-exponential time dependence of magnetization (Fig. 3, Ref. 37) contradicts the predicted dynamics of both DSIM and MMSI models (39). Instead, one intriguing possibility is that the state of $Dy_2Ti_2O_7$ actually represents translationally-invariant many-body localization of the spins (*42,43,44*). It will be fascinating, in this context, to reconsider the absence of magnetic ordering in other frustrated pyrochlores so as to determine if supercooled classical spin liquids occur therein.

**Acknowledgements:** We are grateful to E. Fradkin, B. Gaulin, M. Gingras, S. Grigera, D. Hawthorn, R. Hill, E.-A. Kim, J. Kycia, M.J. Lawler, A.P. Mackenzie, R. Melko, and J. Sethna for very helpful discussions and communications.  This research is funded by the Gordon and Betty Moore Foundation's EPiQS Initiative through Grant GBMF4544 and by the EPSRC Programme Grant 'Topological Protection and NonEquilibrium States in Correlated Electron Systems'.

# FIGURE CAPTIONS

## Figure 1: Signatures of Supercooling in Classic Glass-forming Fluids

A.  Under suitable conditions, a liquid can be cooled through the melting temperature $T_m$ without crystallizing and instead reaches a supercooled liquid state. With further cooling, the heat capacity of the supercooled liquid must eventually decrease from the higher liquid value toward the lower crystalline value; this is necessary for the entropy to remain non-negative at very low temperatures. Supercooled liquids therefore typically have a broad peak in their heat capacity below $T_m$ and a little above the glass transition temperature $T_g$.

B.  In a classic supercooled liquid, diverging microscopic relaxation times $\tau_0(T)$ typically do not follow Arrhenius behavior (dashed line) but instead follow the Vogel-Tammann-Fulcher (VTF) evolution (blue curve; Eqn. 1). Here $T_0$ is a temperature at which the relaxation time diverges to $\infty$ while $D$ characterizes the extent of the super-Arrhenius behavior. By convention, a classic glass is said to form when $\tau_0 > 100$s.

C.  $Re[\varepsilon(\omega, T)]$ and $Im[\varepsilon(\omega, T)]$ of the Havriliak-Negami form (Eqn. 2) of the dielectric function; these are both characteristic of supercooled liquids.

D.  Ultra-slow relaxation in glass-forming liquids occurs with a KWW form (green curve, Eqn. 3) instead of a Debye form (dashed line), as shown here for the dielectric function $\varepsilon$(t).

## Figure 2: Novel Experimental Techniques for Frustrated Magnetism in $Dy_2Ti_2O_7$

A.  The $Dy^{3+}$ moments (black circles) in $Dy_2Ti_2O_7$ are located on a lattice comprised of equilateral corner-sharing tetrahedra. The centers of tetrahedra themselves form a diamond lattice.

B.  An allowed magnetic ground-state configuration for two tetrahedra in $Dy_2T_2iO_7$. Here the crystal field anisotropy causes the moments to point along their local [111] axes, thus forcing them to point toward or away from each tetrahedron center. To minimize magnetic energy, the spin configurations of a tetrahedron of $Dy_2T_2iO_7$ satisfy the spin equivalent of the Bernal-Fowler 'ice rules', with each tetrahedron having two spins pointing toward its center and two spins pointing away from its center.

C. Schematic of a pair of delocalized magnetic monopoles (+ green, - red) representing a sequence of spin flips as shown.

D. Schematic representation of the toroidal geometry of our $Dy_2Ti_2O_7$ sample (yellow) and the superconducting toroidal solenoid (blue). The cryogenic sample environment 30mK<T<4K is indicated by a dashed rectangle. DC current flow in the direction indicated by black arrow produces an azimuthal static magnetic field $B_\phi$ (blue arrows). If a fluid of magnetic-monopoles of both signs exists, the net magnetization current $J_\phi$ (red/green arrows) would be nonzero. Applied AC currents $I_0 Cos(\omega t)$ generate the azimuthal fields $H_\phi Cos(\omega t)$ whose effect is simultaneously detected by measuring the EMF across the STS: $V(\omega, T)$. The dynamical magnetic susceptibility components are then derived from $V_x(\omega, T) = -I_0 \omega L \chi''$, $V_y(\omega, T) = -I_0 \omega L \chi'$, where $L$ is the effective geometrical inductance of the STS.

**Figure 3: Kohlrausch-Williams-Watts Magnetization Transient Dynamics**

A. Typical examples of measured $V(t,T)$ of $Dy_2Ti_2O_7$ torus after the current is switched off in the STS.

B. The KWW collapse of all measured $V(t,T)$ transients from our $Dy_2Ti_2O_7$ torus experiments at the temperatures shown. Clearly, a KWW relaxation model provides an excellent description of the data over the entire range of temperatures at which we could resolve the d$M$/dt signals.

**Figure 4: Havriliak-Negami Magnetic Susceptibility and Vogel-Tammann-Fulcher Microscopic Relaxation Times**

A. Measured frequency and temperature dependence of the real component of the magnetic susceptibility calculated from the EMF component $V_y$ that is 90-degrees out of phase with the applied field.

B. Measured frequency and temperature dependence of the imaginary component of the magnetic susceptibility calculated from the EMF component $V_x$ that is in phase with the applied field.

C. To demonstrate that the Dy$_2$Ti$_2$O$_7$ magnetic susceptibility $\chi(\omega, T)$ is comprehensively described by a Havriliak-Negami form, we define in Eqn. 7 a scaled susceptibility $G(\gamma, \chi)$ depending only upon $\gamma$, $\chi_0$ and $\chi(\omega, T)$. Here we plot Re$[G(\gamma, \chi)]$ vs. $(\omega\tau)^\alpha$ for all our measured $\chi(\omega, T)$ showing a very high precision collapse of all our data to the HN form (shown as a solid line for $[\chi_0 = 1, \gamma = 1, \tau_{HN} = 1, \alpha = 0.91]$). The fit values of γ(T) and α(T) are shown in the insets.

D. Im$[G(\gamma, \chi)]$ plotted versus vs. $(\omega\tau)^\alpha$ for all measured $\chi(\omega, T)$ again showing a high precision collapse of all our data to the HN form (shown as a solid line for $[\chi_0 = 1, \gamma = 1, \tau_{HN} = 1, \alpha = 0.91]$) with identical parameters as in A. The fit values of $\chi_0$ are shown in the inset.

E. Measured microscopic relaxation times $\tau_{HN}$(T) for magnetization in Dy$_2$Ti$_2$O$_7$ spanning the whole temperature range 0.575 K< T < 3 K plotted versus (T-T$_0$)$^{-1}$ where T$_0$ = 242 mK. This demonstrates clearly that magnetic relaxation in Dy$_2$Ti$_2$O$_7$ is governed by the VTF form (Eqn. 1) over almost four orders of magnitude in $\tau_{HN}$.

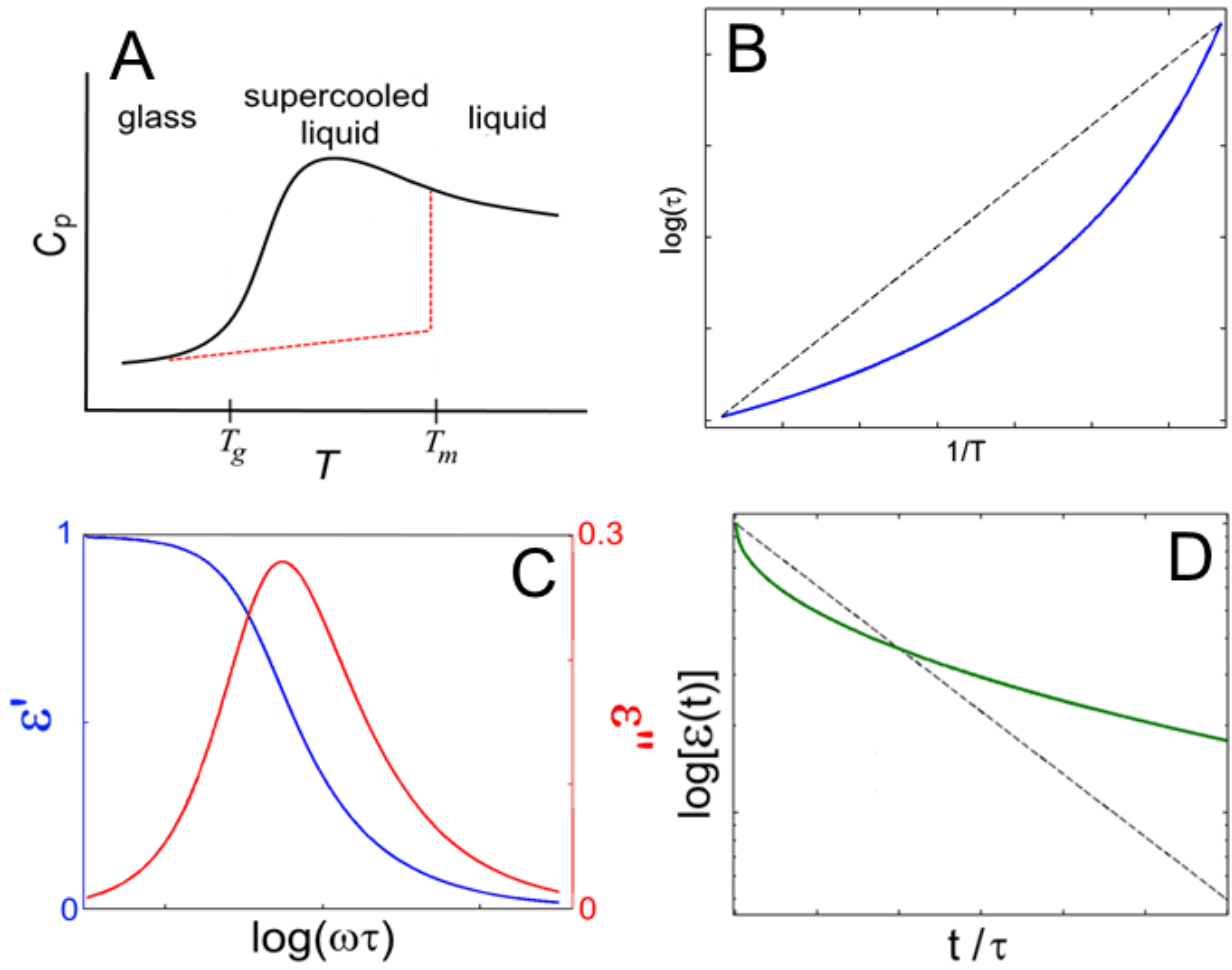

**FIGURE 2**

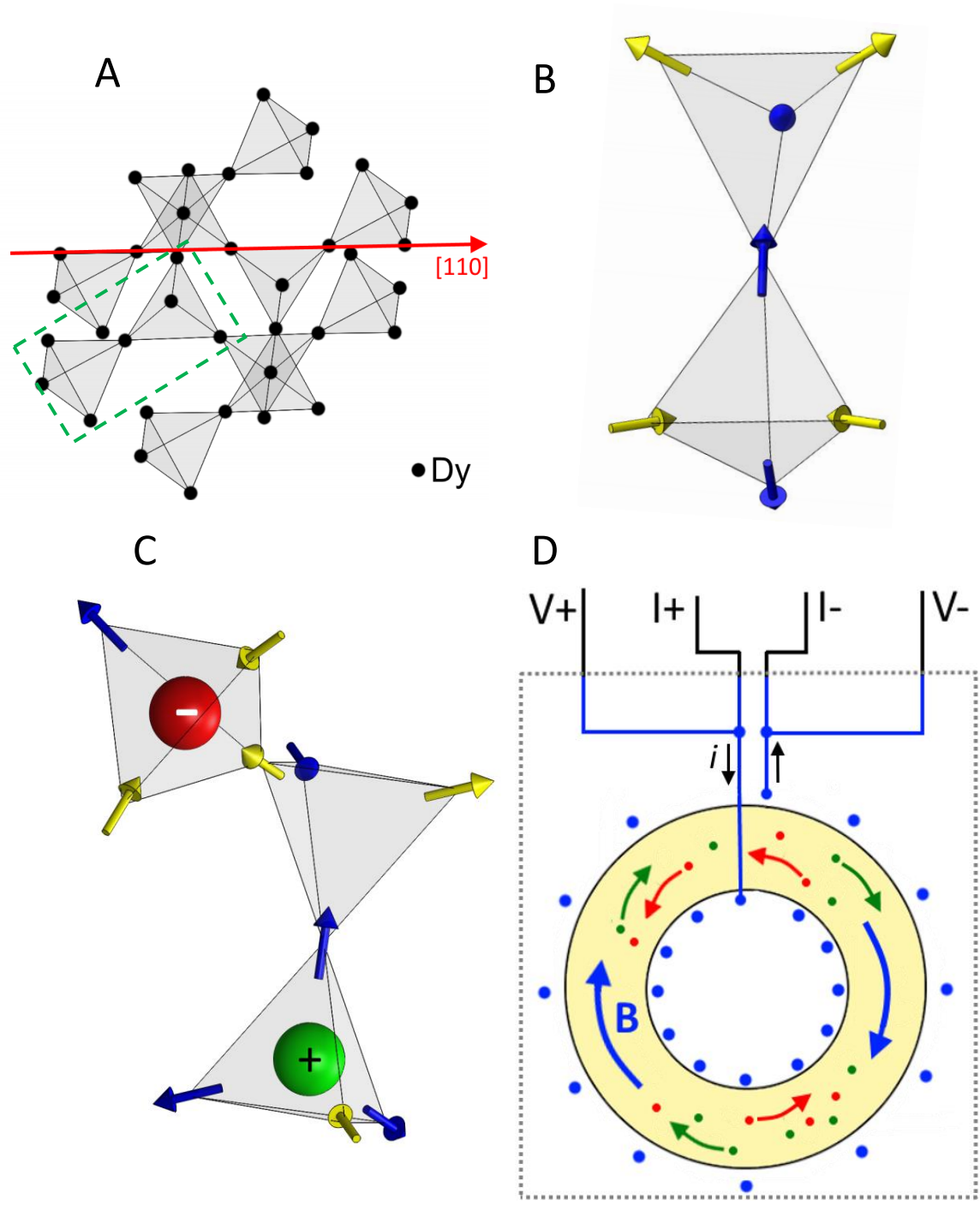

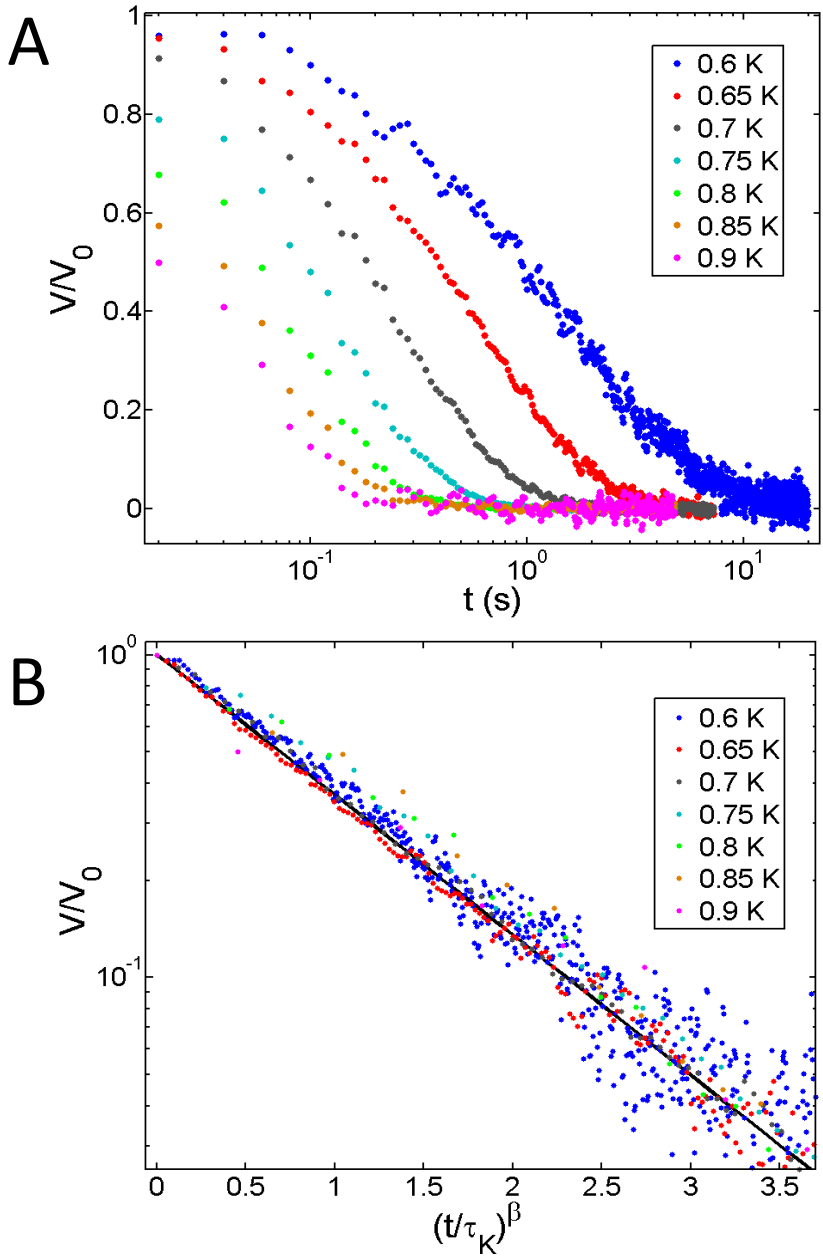

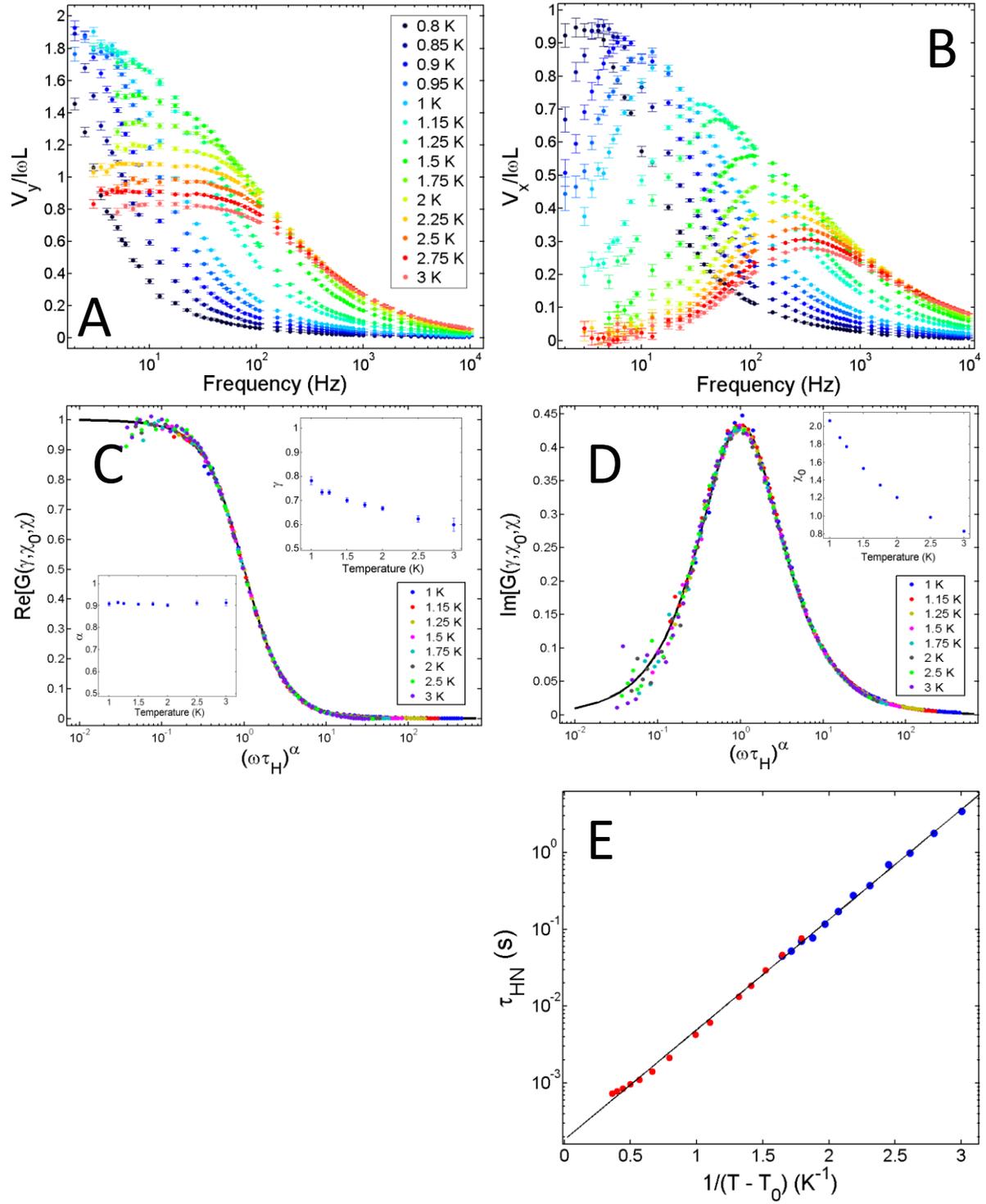

**REFERENCES**

[1] Ediger M, Angell C, Nagel S (1996) Supercooled Liquids and Glasses. *J. Phys. Chem.* 100:13200-13212.

[2] Tarjus G, Kivelson S, Nussinov Z, Viot P (2005) The frustration-based approach of supercooled liquids and the glass transition: a review and critical assessment. *J. Phys: Condens. Matt.* 17:R1143-R1182.

[3] Cavagna A (2009) Supercooled liquids for pedestrians. *Physics Reports* 476:51-124.

[4] Bohmer R, Ngai K, Angell C, Plazek D (1993) Nonexponential relaxations in strong and fragile glass formers. *J. Chem. Phys.* 99:4201-4209.

[5] Havriliak S, Negami S (1967) A Complex Plane Representation of Dielectric and Mechanical Relaxation Processes in Some Polymers. *Polymer* 8:161-210.

[6] Havriliak Jr. S, Havriliak S (1994) Results from an unbiased analysis of nearly 1000 sets of relaxation data. *J. Non-Cryst. Solids* 172-174:297-310.

[7] Kohlrausch R (1854) Theorie des elektrischen ruckstandes in der leidner flasche. *Annelen der Physik und Chemie (Poggendorff)* 91:179-213.

[8] Alvarez F, Alegría A, Colmenero J (1991) Relation between the time-domain Kohlrausch-Williams-Watts and frequency-domain Havriliak-Negami relaxation functions. *Phys. Rev. B* 44:7306-7312.

[9] Harris M., Bramwell S, Holdsworth P, Champion J (1998) Liquid-Gas Critical Behavior in a Frustrated Pyrochlore Ferromagnet. *Phys. Rev. Lett.* 81:4496-4499.

[10] (2011) *Introduction to Frustrated Magnetism: Materials, Experiments, Theory*, eds Lacroix C, Mendels P, Mila F (Springer Series in Solid-State Sciences, 2011).

[11] Gardner J, Gingras M, Greedan E (2010) Magnetic pyrochlore oxides. *Rev. Mod. Phys.* 82:53-107.

[12] Balents L (2010) Spin liquids in frustrated magnets. *Nature* 464:199-208.

[13] Castelnovo C, Moessner R, Sondhi S (2012) Spin Ice, Fractionalization, and Topological Order. *Annu. Rev. Condens. Matter Phys.* 3:35-55.

[14] Gingras M, McClarty P (2014) Quantum spin ice: a search for gapless quantum spin liquids in pyrochlore magnets. *Rep. Prog. Phys.* 77:056501.

15  Rosenkranz S *et al* (2000) Crystal-field interaction in the pyrochlore magnet Ho2Ti2O7. *Journal of Applied Physics* 87:5914-5916.

16  Fukazawa H, Melko R, Higashinaka R, Maeno Y, Gingras M (2002) Magnetic anisotropy of the spin-ice compound Dy2Ti2O7. *Phys. Rev. B* 65:054410.

17  Dunsiger S *et al.* (2011) Magnetic Excitations without Monopole Signatures using Muon Spin Rotation. *Phys. Rev. Lett.* 107:207207.

18  Klemke B *et al.* (2011) Thermal Relaxation and Heat Transport in the Spin Ice Material Dy2Ti2O7. *J. Low Temp. Phys.* 163:345-369.

19  Higashinaka R, Fukazawa H, Yanagishima D, Maeno Y (2002) Specific heat of Dy2Ti2O7 in magnetic fields: comparison between single-crytalline and polycrystalline data. *J. Phys. Chem. Solids*. 63:1043-1046.

20  Pomaranski D *et al.* (2013) Absence of Pauling's residual entropy in thermally equilibrated Dy2Ti2O7. *Nat. Phys.* 9:353–356.

21  Ramirez A, Hayashi A, Cava R, Siddharthan R (1999) Zero-point entropy in 'spin ice'. *Nature* 399:333-335.

22  Morris D *et al.* (2009) Dirac Strings and Magnetic Monopoles in the Spin Ice Dy2Ti2O7. *Science* 326:411-414.

23  den Hertog B, Gingras M (2000) Dipolar Interactions and the Origin of Spin Ice in Ising Pyrochlore Magnets. *Phys. Rev. Lett.* 84:3430.

24   Bramwell S, Gingras M (2001) Spin Ice State in Frustrated Magnetic Pyrochlore Materials. *Science* 294:1495-1501.

25  Melko R, Gingras M (2004) Monte Carlo studies of the dipolar spin ice model. *J. Phys. Condens. Matter*. 16:R1277–R1319.

26  Castelnovo C, Moessner R, Sondhi S (2008) Magnetic monopoles in spin ice. *Nature*. 451:42-45.

27 Levin Y (2002) Electrostatic correlations: from plasma to biology. *Rep. Prog. Phys* 65:1577-1632.

28 Castelnovo C, Moessner R, Sondhi S (2011) Debye-Hückel theory for spin ice at low temperature. *Phys. Rev. B* 84:144435.

29 Yavors'kii T, Fennell T, Gingras M, Bramwell S (2008) Dy2Ti2O7 Spin Ice: A Test Case for Emergent Clusters in a Frustrated Magnet. *Phys. Rev. Lett.* 101:037204.

[30] Snyder J *et al.* (2004) Low-temperature spin freezing in the Dy2Ti2O7 spin ice. *Phys. Rev. B* 69:064414.

[31] Matsuhira K *et al.* (2011) Spin Dynamics at Very Low Temperature in Spin Ice Dy2Ti2O7. *J. Phys. Soc. Japan* 80:123711.

[32] Bovo L, Bloxsom J, Prabhakaran D, Aeppli G, Bramwell S (2013) Brownian motion and quantum dynamics of magnetic monopoles in spin ice. *Nat. Comm.* 4:1535-1542.

[33] Yaraskavitch L *et al.* (2012) Spin dynamics in the frozen state of the dipolar spin ice Dy2Ti2O7. *Phys. Rev. B* 85:020410.

[34] Takatsu H *et al.* (2013) AC Susceptibility of the Dipolar Spin Ice Dy2Ti2O7: Experiments and Monte Carlo Simulations. *J. Phys. Soc. Japan* 82:104710.

[35] Giblin S, Bramwell S, Holdsworth P, Prabhakaran D, Terry I (2011) Creation and measurement of long-lived magnetic monopole currents in spin ice. *Nat. Phys.*, 7:252-258.

[36] Orendáĉ M *et al.* (2007) Magnetocaloric study of spin relaxation in dipolar spin ice Dy2Ti2O7. *Phys. Rev. B* 75:104425.

[37] Revell H *et al.* (2012) Evidence of impurity and boundary effects on magnetic monopole dynamics in spin ice. *Nat. Phys* 9:34-37.

[38] Jaubert L, Holdsworth P (2009) Signature of magnetic monopole and Dirac string dynamics in spin ice. *Nat. Phys* 5:258-261.

[39] Jaubert L, Holdsworth P (2011) Magnetic monopole dynamics in spin ice. *J. Phys.: Condens. Matt.* 23:164222.

[40] Cépas O, Canals B (2012) Heterogeneous freezing in a geometrically frustrated spin model without disorder: Spontaneous generation of two timescales. *Phys. Rev. B* 86:024434.

[41] Z. Nussinov, C. D. Batista, B. Normand, and S. A. Trugman (2007) High-dimensional fractionalization and spinon deconfinement in pyrochlore antiferromagnets Phys. Rev. B 75: 094411

[42] De Roeck W, F. Huveneers (2014) Scenario for delocalization in translation-invariant systems. *Phys. Rev. B* 90:165137.

[43] Yao N, Laumann C, Cirac J, Lukin M, Moore J (2014) Quasi Many-body Localization in Translation-Invariant Systems. *arXiv* 1410.7407

[44] Schiulaz M, Silva A, Müller M, (2014) Dynamics in many-body localized quantum systems without disorder. *arXiv* 1410.4690.

# A Supercooled Spin Liquid State
# in the Frustrated Pyrochlore $Dy_2Ti_2O_7$

Ethan R. Kassner, Azar B. Eyvazov, Benjamin Pichler, Timothy J. S. Munsie,
Hanna A. Dabkowska, Graeme M. Luke & J.C. Séamus Davis

## S1. Toroidal $Dy_2Ti_2O_7$ sample and Toroidal Superconducting Sensor Coil

The $Dy_2Ti_2O_7$ samples used in this project were synthesized by the Graeme Luke Group at McMaster University using an optical floating zone furnace. The original sample boule was grown at a rate of $\sim$ 7mm/hr in $O_2$ gas under 2 atm of pressure, and was subsequently cut into disks of diameter $\sim$ 6 mm and thickness $\sim$ 1 mm. X-ray diffraction on the resultant crystal was sharp and showed no signs of twinning or the presence of multiple grains. Performance of Rietveld refinement on the diffraction data yields a unit cell lattice constant of 10.129 Å; this implies a maximum possible level of "spin stuffing" (substitution of $Dy^{3+}$ ions on $Ti^{4+}$ sites) of $\approx$ 2.9% and a most likely spin stuffing fraction < 1%.

To create boundary-free conditions for our measurements of magnetization dynamics, we pierced holes of diameter $\sim$ 2.5 mm through each $Dy_2Ti_2O_7$ disk using diamond-tipped drill bits. Fig. S1A shows two typical samples after the completion of the drilling process. We also manufactured equivalent tori from Stycast 1266 for control measurements; these were cut and drilled to match the geometrical details of our $Dy_2Ti_2O_7$ samples. Because Stycast 1266 has negligible magnetic activity in our parameter space, it functioned as a good material for control tests; we observed no significant temperature-dependent magnetic signals in this epoxy.

CuNi-clad NbTi wire (thickness 0.1 mm) was wound around each toroidal $Dy_2Ti_2O_7$ sample creating a superconducting toroidal solenoid (STS) (Fig S1B). Lakeshore varnish (VGE-7031) was used to mount our samples on a stage connected to the mixing chamber of a dilution refrigerator and performed measurements at temperatures from 30 mK up to 3

K. Sample temperature was measured with a Lakeshore RuO$_2$ (Rox$^{TM}$) thermometer, and temperature was controlled using a Lakeshore 370 AC Resistance Bridge.

## S2. AC magnetization dynamics experiments

For AC measurements four-probe I-V circuit configuration (Fig 2D in the main text) was utilized . A lock-in amplifier was used to apply AC currents (of up to 30 mA) to the STS and to simultaneously measure the STS EMF. We begin our AC analysis by writing the STS EMF in terms of the changing magnetic flux:

$$V = -\frac{d\Phi}{dt} = -NA\frac{dB}{dt} = -NA\mu_0\frac{d}{dt}(H + M)$$

where $N$ is the number of coil turns in the STS and $A$ is the effective coil cross-sectional area. Using the standard definition of the frequency-dependent magnetic volume susceptibility, $M(\omega) = \chi(\omega)H(\omega)$, and the applied magnetic field $H = nI\exp(i\omega t)$, one can calculate the amplitude of the STS EMF:

$$V(\omega) = -\mu_0 NAnIi\omega\big(1 + \chi(\omega)\big) = -iI\omega L(1 + \chi(\omega)) \qquad \text{(S1)}$$

where $N$ is total number of STS turns, $n$ is the turns/length in the STS, $A$ is the effective cross-sectional area of the coil, $I$ is the applied current, and $L$ is the effective geometric inductance of the STS. We specifically measured the EMF components that were in phase ("X") and 90 degrees out of phase ("Y") with the applied current: $V = V_x + iV_y$. Experiments on our nonmagnetic Stycast 1266 control sample, which used the same circuitry as our Dy$_2$Ti$_2$O$_7$ measurements, showed no temperature dependence; all signals in the Stycast coils were due simply to standard circuit responses to a changing magnetic field. These observations confirmed that the circuit geometry and conductance in the area of the experiment had no significant variation in our measured temperature range. Any temperature-dependent changes in the frequency-dependent STS EMF are due to changes in the magnetization dynamics of the Dy$_2$Ti$_2$O$_7$ samples.

Since the microscopic relaxation times of $Dy_2Ti_2O_7$ are sensitive to its temperature (*1*), it was essential to verify that our samples and measurements stabilized at each set temperature. We determined thermalization times by performing tests in which we took lock-in amplifier readings for several hours after temperature and frequency changes; Figure S2 shows typical results of these measurements. After frequency changes the EMF readings settled almost immediately (within seconds) to stable long-time values, while the readings reached stable values after less than 10 minutes following temperature changes. To accommodate this thermalization time, we waited at least 15 minutes after temperature changes before recording data in both the AC and DC experiments.

We measured the geometric inductance $L$ of our STS systems by measuring $V_y$ at frequencies in the range 40-80 kHz at T=30 mK; these frequencies are well above the frequency range in which we observed significant $Dy_2Ti_2O_7$ dynamics, so for these measurements the STS EMF is dominated by the vacuum inductor signal $V_y = I\omega L$. From the linear frequency dependence of $V_y$ we have found that our typical STS inductances are 1-2 μH.

Figs S3A and S3B show the real and imaginary parts, respectively, of typical STS EMF data at 500 mK and 900 mK, with 50-mK data treated as background and subtracted. There is a clear difference in the EMFs measured at 900 mK and 50 mK; we use this temperature dependence of the STS signals to study the frequency-dependent magnetization dynamics of $Dy_2Ti_2O_7$ (see below). At 500 mK, however, the STS EMF is indistinguishable from the EMF at 50 mK. Previous susceptibility measurements (2) indicate that $Dy_2Ti_2O_7$ has negligible magnetic activity in our frequency range (2-1000 Hz) at temperatures as low as several hundred mK; we therefore assume that the data at 50 mK is due completely to non-$Dy_2Ti_2O_7$ sources, and we conclude that all data in our parameter space taken at $T \leq 500$ mK are due to non- $Dy_2Ti_2O_7$ background sources.

For experiments at higher temperatures, we subtracted the 500-mK data to isolate $Dy_2Ti_2O_7$ contributions to the STS EMF:

$$\Delta V(\omega, T) = V(\omega, T) - V(\omega, 500 \; mK) = \Delta V_x(\omega, T) + i\Delta V_y(\omega, T)$$
$$= -iI\omega L(\chi' - i\chi'')$$

where we have used the conventional notation for the complex magnetic volume susceptibility $\chi = \chi' - i\chi''$. Therefore, the quadrature EMF components can be written as

$$\Delta V_x = -I\omega L\chi'' \qquad\qquad (S2)$$
$$\Delta V_y = -I\omega L\chi'$$

The HN form for a magnetic susceptibility is given by

$$\chi(\omega, T) = \chi_\infty + \frac{\chi_0}{[1 + (i\omega\tau_{HN})^\alpha]^\gamma} \qquad\qquad (S3)$$

where $\chi_0$ is a relaxation amplitude, $\tau_{HN}$ is a characteristic relaxation time, and the exponents $\alpha$ and $\gamma$ describe the spread and asymmetry, respectively, of the relaxation in frequency space. $\chi_\infty$ gives the relaxation in the $\omega \to \infty$ limit. This model is a generalization of the simplest complex relaxation models: Debye relaxation has $\alpha = 1$, $\gamma = 1$; Cole-Cole relaxation has $\gamma = 1$; and Davidson-Cole relaxation has $\alpha = 1$. Figures S4A and S4B show the quadrature susceptibility components at two representative temperatures, along with fits to Debye and HN forms. Debye relaxation is clearly an inadequate description of our data; one needs the full set of HN parameters to reproduce the observed magnetization dynamics.

Figures S5A and S5B show the full set of real and imaginary parts of the susceptibility using and Equation (S2). These experiments were performed between 0.8 K and 3 K. We performed simultaneous fits of $\chi'(\omega)$ and $\chi''(\omega)$ to the HN function using a least squares method that minimized the cumulative residuals of the real and imaginary susceptibility components. The solid lines in Figs S5A and B show the results of these fits; our observations are clearly described very well by HN relaxation. The small residuals ($\chi'(\omega, T) - \chi'_{HN}(\omega, T)$, inset of Fig. S5A; $\chi''(\omega, T) - \chi''_{HN}(\omega, T)$, inset of Fig. S5B) further illustrate the quality of these fits; throughout all of our parameter space the residuals are a few percent or less of the signal size. HN spectra are generally found in supercooled liquids. The temperature dependencies of the HN parameters $\chi_0$ (Fig. S6A), $\chi_\infty$ (Fig. S6B), $\alpha$ (Fig. S6C), and $\gamma$ (Fig. S6D) are shown in Fig. S6; $\tau_{HN}$ is shown along with the relaxation times obtained from our DC experiments in Figure 4E of the main text.

We can further show the broad applicability of HN relaxation to $Dy_2Ti_2O_7$ dynamics by finding scaled HN variables that collapse the $T$-dependent $\chi'$ and $\chi''$ data onto single curves. To do this, we start with the basic HN form and separate the exponents $\alpha$ and $\gamma$:

$$\chi(\omega, T) = \chi' - i\chi'' = \frac{\chi_0}{[1 + (i\omega\tau_{HN})^\alpha]^\gamma}$$

$$\left(\frac{\chi' - i\chi''}{\chi_0}\right)^{1/\gamma} = \frac{1}{1 + (i\omega\tau_{HN})^\alpha} \tag{S4}$$

where we ignore the effects of $\chi_\infty$ since it is much smaller than $\chi_0$ (Figure S6). Using complex algebra identities, we can write the LHS of Equation (S4) as

$$G(\gamma, \chi_0, \chi) \equiv \left(\frac{\chi'^2 + \chi''^2}{\chi_0^2}\right)^{1/2\gamma} \left(\cos\left(\frac{1}{\gamma}\arctan\frac{\chi''}{\chi'}\right) - i\,\sin\left(\frac{1}{\gamma}\arctan\frac{\chi''}{\chi'}\right)\right) \tag{S5}$$

and we can write the RHS as

$$H(\alpha, \omega\tau_{HN}) \equiv \frac{1 + (\omega\tau_{HN})^\alpha(\cos\frac{\pi\alpha}{2} - i\sin\frac{\pi\alpha}{2})}{1 + 2(\omega\tau_{HN})^\alpha\cos\frac{\pi\alpha}{2} + (\omega\tau_{HN})^{2\alpha}} \tag{S6}$$

The real and imaginary components of $G$ give us effective scaled variables for the real and imaginary susceptibility components, respectively:

$$\mathrm{Re}[G(\gamma, \chi_0, \chi)] = \chi_0^{-1/\gamma}[(\chi')^2 + (\chi'')^2]^{\frac{1}{2\gamma}}\cos\left[\frac{1}{\gamma}\arctan\left(\frac{\chi''}{\chi'}\right)\right]$$

$$\mathrm{Im}[G(\gamma, \chi_0, \chi)] = \chi_0^{-1/\gamma}[(\chi')^2 + (\chi'')^2]^{\frac{1}{2\gamma}}\sin\left[\frac{1}{\gamma}\arctan\left(\frac{\chi''}{\chi'}\right)\right]$$

To complete the data collapse procedure, we need to choose a suitable scaled frequency; the best choice is $x \equiv (\omega\tau_{HN})^\alpha$. Using this definition we can rewrite the HN relaxation as

$$G(\gamma, \chi_0, \chi) = \frac{1 + x(\cos\frac{\pi\alpha}{2} - i\sin\frac{\pi\alpha}{2})}{1 + 2x\cos\frac{\pi\alpha}{2} + x^2} \qquad (S7)$$

α varies weakly with temperature (Fig. S6C), so with these definitions we achieve excellent universal collapse of our AC data (Figs. 4C and 4D in the main text). The characteristic curve of this collapse is a Cole-Cole curve with $\alpha = 0.91$.

## S3. Time-domain experiments

During time domain experiments we used a 4-probe superconducting I-V circuit configuration to apply current to and measure the EMF across our STSs (Fig. 2D in the main text). For these studies we applied DC currents as high as 25 mA using an Agilent 33210A Arbitrary Waveform Generator to create a current supply; we simultaneously measured the STS EMF, *V(t)*, with a Keithley 2182A Nanovoltmeter. As shown in Fig. S7A, the current application protocol is as follows (green dashed lines): (1) apply current in one direction for a time interval *t*, (2) turn off the current for an identical interval *t*, (3) apply current in the opposite direction for *t*, (4) turn off the current for *t* and then return to step 1. We typically repeated this sequence hundreds of times to improve our signal-to-noise ratio and achieve robust fits to the data. Fig S7A depicts a typical complete measurement sequence at three experimental temperatures. The STS *V(t)* was measured every 20 milliseconds throughout the experiment, and we observed that 20 seconds is sufficient for the induced EMF to drop below our noise level ~1 nV at all temperatures above 550 mK. Our ability to resolve decay curves at higher temperatures was limited by the time resolution of the nanovoltmeter.

Our *V(t)* are direct measurements of changes in the sample magnetization density *M* over time:

$$V = -N\frac{d\Phi}{dt} = -\mu_0 NA \frac{d(M+H)}{dt} \tag{S8}$$

$$= -\mu_0 NA \frac{dM}{dt} - \mu_0 NA \frac{dH}{dt}$$

where $N$ is the number of coil turns in the STS and $A$ is the effective STS cross-sectional area. The second term, which describes the STS response to changes in the applied field itself, is present even when the STS encloses vacuum. Since the only change in $H$ occurs when we turn the applied current on or off, the contributions of this term are limited to very short times ($< \sim 100$ ms) after current changes. For our fits we examined data taken $\geq$ 200 ms after current changes, when only the first term in Equation (S8) contributes.

We can relate the measured EMF to the time-dependent magnetic volume susceptibility $\chi(t)$ by considering the STS as having inductance $L$ with $I$ as the field-generating current. Taking the time derivative of $M(t)$ and inserting it into Equation (S8) gives the final expression for the long-time STS EMF:

$$V = IL\chi(t) \tag{S9}$$

Fig S7B depicts the typical measured EMF (symbols) generated for $t \geq$ 200 ms after turning off the STS current at temperatures from 575 mK up to 900 mK; at these times the EMF is given by Equation (S9). The STS EMF shows slower-than-exponential decay, and the decay times increase dramatically with decreasing temperature. The lines in Fig. S6B are fits to the Kohlrausch-Williams-Watts (KWW) form

$$V(t) = V_0 \exp\left(-\left(\frac{t}{\tau_{KWW}}\right)^{\beta}\right) \tag{S10}$$

where $\tau_{KWW}$ is a relaxation time and $\beta$ is a stretching exponent. Here $\beta$ = 1 corresponds to standard Debye relaxation, while $\beta$ < 1 indicates the presence of a more complex landscape of energy barriers and dynamics. The KWW fits to our measured $V(t)$ were performed using a least squares method.

Figure S7B shows that the KWW form describes the time-domain magnetization dynamics very well. If there is truly universal applicability of the KWW function to our data, we should be able to define scaled variables such that data from all temperatures collapse onto a single function. We achieve this by defining a scaled time $x \equiv (\frac{t}{\tau_{KWW}})^\beta$ and plotting $v \equiv \log \frac{V(t)}{V_0}$ against this new parameter. From Equation (S10), relaxation with a KWW form should collapse onto a simple exponential given by $v = \exp(-x)$, which is a straight line on a log-linear plot; indeed this is what we find (Fig. 3B in the main text).

The inset to S7B shows residuals for the KWW fits, i.e. $V(t) - V_{KWW}(t)$. The magnitude and time independence of these residuals indicate that the KWW model is an excellent description of all our $V(t)$ data. Figure S8 shows temperature dependence of the exponent $\beta$; this exponent shows a weak increasing trend, but in most of the temperature range we find $\beta \approx 0.75$, indicating the presence of complex relaxation dynamics in $Dy_2Ti_2O_7$. The temperature dependence of $\tau_{KWW}$ is shown in Fig. 4C of the main text, as is the behavior of the relaxation time obtained from our AC experiment (see below).

## S4. Super-Arrhenius Diverging Microscopic Relaxation Times

The temperature dependence of the observed relaxation times differs substantially from the standard Arrhenius form $\tau = A\exp(\Delta/k_B T)$. Supercooled dielectric liquids are known to possess dynamical quantities, such as the viscosity and microscopic dielectric relaxation time, that exhibit a super-Arrhenius temperature variation described by the Vogel-Tammann-Fulcher (VTF) form (3,4,5):

$$\tau_0(T) = A\exp\left(\frac{DT_0}{T - T_0}\right) \tag{S11}$$

Here $D$ characterizes the extent of departure from an Arrhenius form (often referred to as the "fragility" of the liquid), and $T_0$ is a temperature at which the relaxation time diverges.

From experiments we know that $T_0$ gives an effective lower bound on the temperature at which a glass-forming liquid must transition into the glass state or crystallize (*5*).

      To apply this formalism to our $Dy_2Ti_2O_7$ measurements, we must present the results from our separate DC and AC measurements in a unified manner. We have done this by converting $\tau_{KWW}$ from DC measurements into $\tau_{HN}$ using the relations in Equation 4 of the main text; the complete temperature dependence of the relaxation times is plotted in Fig 4E in the main text and in Fig S9. We performed AC and DC experiments at overlapping temperatures of 800 mK – 850 mK to make sure that the relaxation times obtained by these two independent techniques give a consistent description of $Dy_2Ti_2O_7$ magnetization dynamics. Fig S9 shows that the microscopic relaxation time in $Dy_2Ti_2O_7$ varies smoothly throughout our entire temperature range, even when we cross over from our AC results to DC results. Results for $\tau(T)$ from AC and DC measurements are equal within the uncertainty at the overlap temperatures. Therefore, they yield a single smooth unified description of $Dy_2Ti_2O_7$ dynamics from 575 mK up to 3 K. $\tau(T)$ exhibits super-Arrhenius behavior that is captured by the VTF form in Equation (S11).

# FIGURE CAPTIONS

**Figure S1: Samples**

A. Two typical drilled $Dy_2Ti_2O_7$ samples; these tori have OD $\sim$ 6 mm, ID $\sim$ 2.5 mm, and thickness $\sim$ 1 mm.

B. DTO samples after being wound with fine 0.1-mm-thick, CuNi-clad NbTi wire to create a Superconducting Toroidal Solenoid (STS). The STS allows us to measure $Dy_2Ti_2O_7$ magnetization dynamics in a situation where any putative magnetic fluid does not encounter crystal boundaries. It also minimizes demagnetization effects that are present in cylindrical or rod geometries.

**Figure S2: Thermalization tests**

A. Time series of the in-phase EMF component $V_x$ after frequency changes and temperature changes. The EMF settles to stable values seconds after frequency changes and < 10 minutes after temperature changes.

B. Time series of the out-of-phase EMF component $V_y$ after frequency changes and temperature changes. The relaxation behavior of this component is similar to what was observed for the in-phase component.

**Figure S3: AC background analysis**

A. Typical changes in the in-phase EMF $\Delta V_x$ from the EMF measured at 50 mK. The $T$-dependent signal is negligible at 500 mK, while we see clear EMF changes at 900 mK. Since the superconducting circuit geometry and conductance characteristics do not significantly change in this temperature range, we conclude that the data at 500 mK can be treated as a background signal, and that changes in the EMF at higher temperatures are due to $Dy_2Ti_2O_7$ magnetization dynamics.

B. Typical changes in the out-of-phase EMF quadrature component $\Delta V_y$. Once again we see no significant signal changes at 500 mK, while measurements at higher temperatures find clear contributions from $Dy_2Ti_2O_7$ magnetization dynamics.

**Figure S4: Necessity of the HN form for Magnetic Susceptibility**

A. Real part of the AC susceptibility (symbols) measured at 1.15 K and 1.75 K, with fits to a Debye relaxation form (dashed lines) and HN relaxation form (solid lines). Debye relaxation clearly cannot describe the data, while HN relaxation provides excellent fits to the data.

B. Imaginary part of the measured AC susceptibility (symbols) at 1.15 K and 1.75 K, with fits to a Debye relaxation form (dashed lines) and HN relaxation form (solid lines). As in (A), HN relaxation provides a far superior empirical descriptuon of the data.

**Figure S5: AC susceptibility measurements**

A. Real part of the AC susceptibility (symbols) calculated from the out-of-phase EMF component $V_y$ between 0.8 K and 3 K. Solid lines are the fits to the HN relaxation model. Inset: Residuals $\chi'(\omega) - \chi'_{HN}(\omega)$ for the HN fit; throughout our parameter space the residuals are a few percent or less of the signal size, indicating that HN susceptibility provides an excellent global description of our observations.

B. Imaginary part of the AC susceptibility (symbols) calculated from the in-phase EMF component $V_x$ between 0.8 K and 3 K. Once again HN fits to the data (lines) are excellent throughout the temperature range. Inset: Residuals $\chi''(\omega) - \chi''_{HN}(\omega)$ for the HN fit; again the residuals are small and only a few percent or less of the signal size.

**Figure S6: AC susceptibility fitting parameters.**

A. Temperature dependence of the HN relaxation amplitude $\chi_0$, which gives the HN susceptibility in the $\omega \to 0$ limit.

B. Temperature dependence of the limiting susceptibility $\chi_\infty$, which gives the susceptibility in the $\omega \to \infty$ limit. $\chi_\infty$ is less than 0.5% of $\chi_0$ throughout our temperature range, so it does not contribute significantly to DTO dynamics in our frequency bandwidth.

C. Temperature dependence of the HN exponent α, which describes the overall width of the $\chi''$ peak in frequency space. α varies only weakly with temperature, allowing us to describe data at all temperatures with a universal functional form (Section S5).

D. Temperature dependence of the HN exponent γ, which describes the asymmetry of the $\chi''$ peak in frequency space. γ decreases substantially with increasing temperature, indicating that $Dy_2Ti_2O_7$ relaxation dynamics become increasingly skewed toward higher frequencies at higher temperatures.

**Figure S7: Time domain relaxation dynamics**

A. Typical results of a complete DC measurement sequence at 0.6 K, 0.7 K, and 0.8 K (symbols). A 25-mA current (green dashed line) is applied at t=0 s, turned off at t=20 s, applied in the opposite direction at t=40 s, and turned off again at t=60s. This current sequence is repeated hundreds of times to achieve good statistical noise levels. A very sharp signal peak and fast decay occur within ∼ 100 ms (independent of temperature) of current changes; after this the signals settle into smooth decays. The initial signal spike is dominated by instrumental/circuit effects, while the subsequent smooth decay is dominated by $Dy_2Ti_2O_7$ dynamics.

B. EMF decays (symbols) measured across the STS after turning off the applied current. We can describe these decays very well with a KWW relaxation model (solid lines). Inset: Residuals of KWW fits to the data are a few percent or less of the signal size throughout our parameter range, demonstrating the very good quality and universality of these fits.

**Figure S8: KWW stretching exponent**

Temperature dependence of the KWW stretching exponent β from fits to the data in Fig. S2.

**Figure S9: The complete set of $Dy_2Ti_2O_7$ relaxation times**

Relaxation times as determined from HN fits to the AC susceptibility (red circles) and as calculated from time domain fits using Equation (4) in the main text (blue circles). The measured divergence of the relaxation times with decreasing temperature cannot be captured by an Arrhenius fit (dashed line). By contrast, the VTF form (solid line) for $\tau(T)$ describes the phenomena excellently.

# FIGURES

## FIGURE S1

A

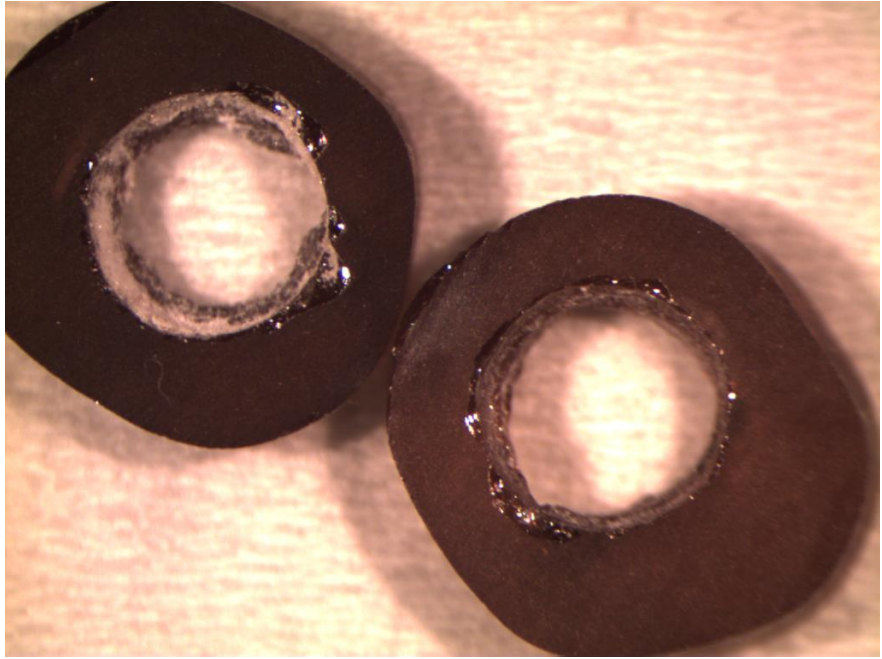

B

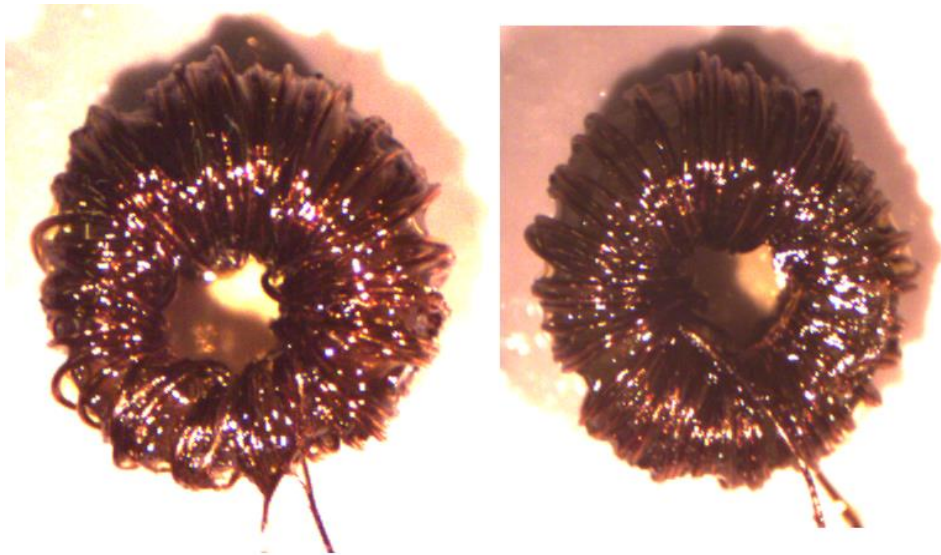

**FIGURE S2**

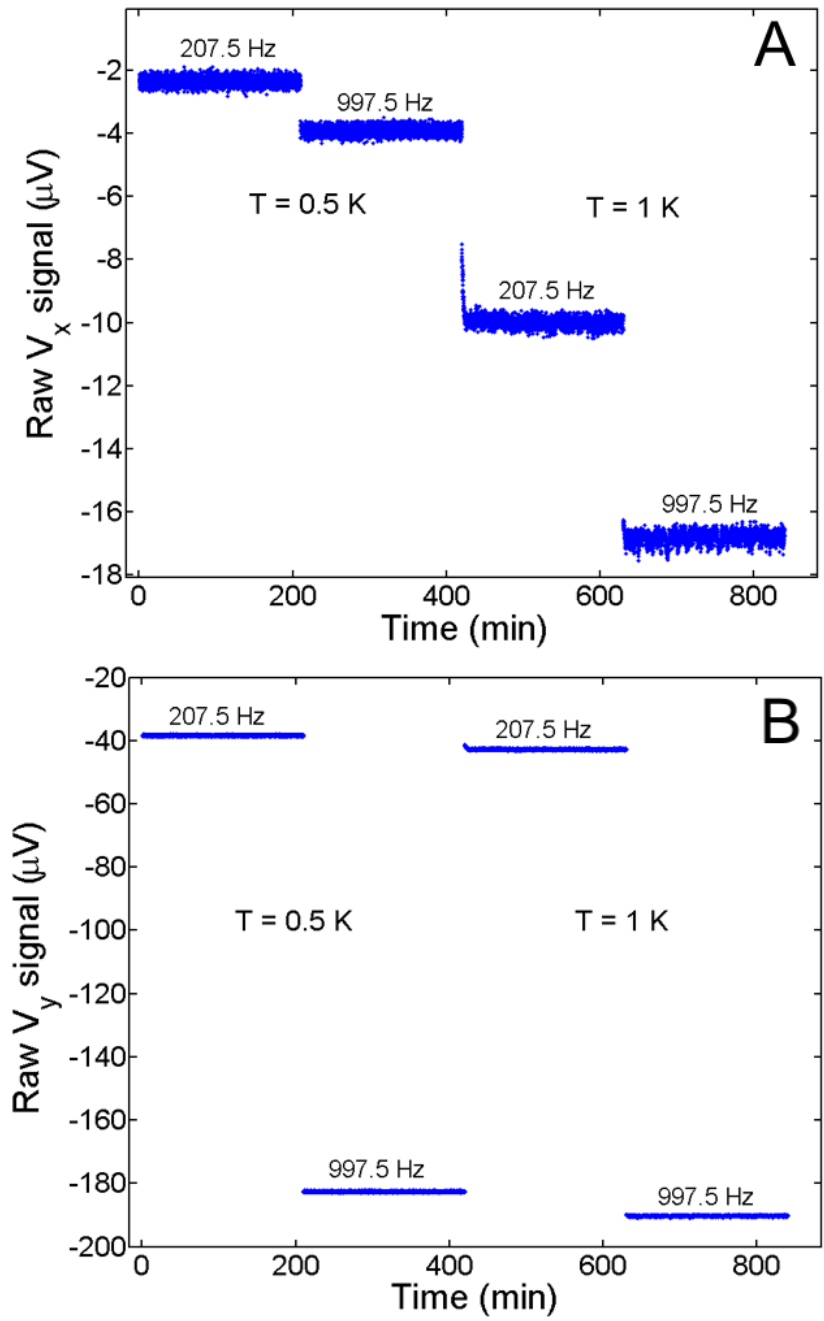

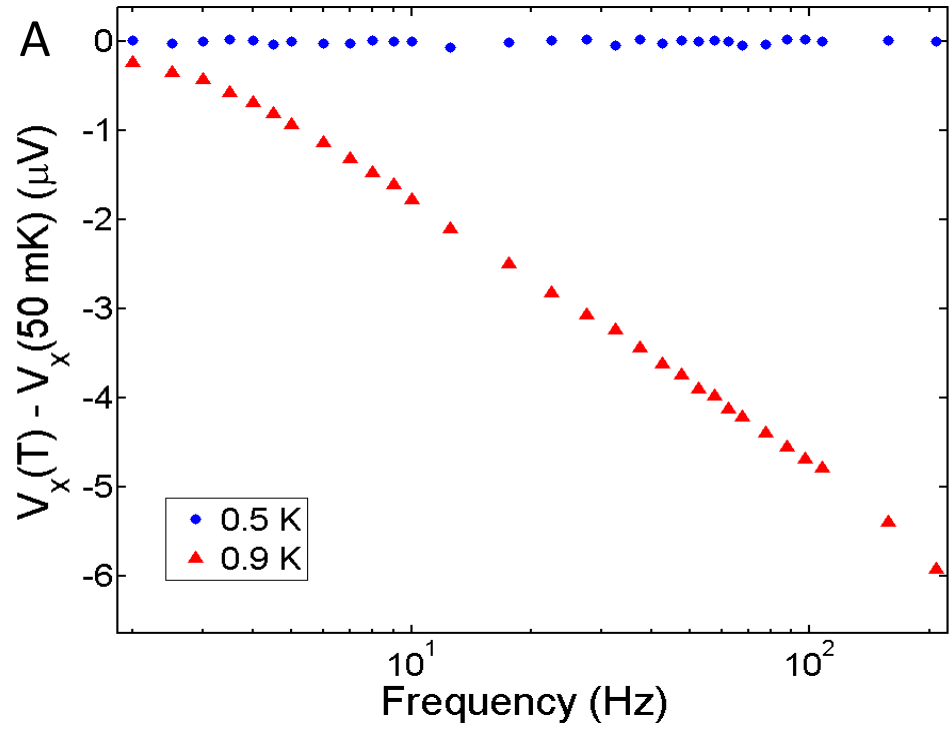

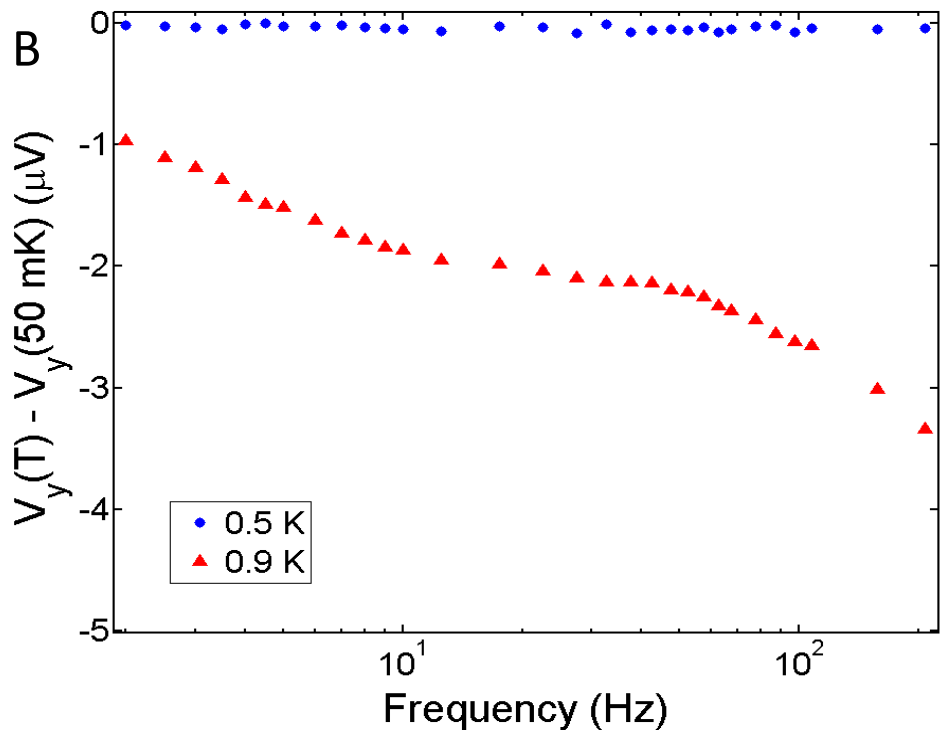

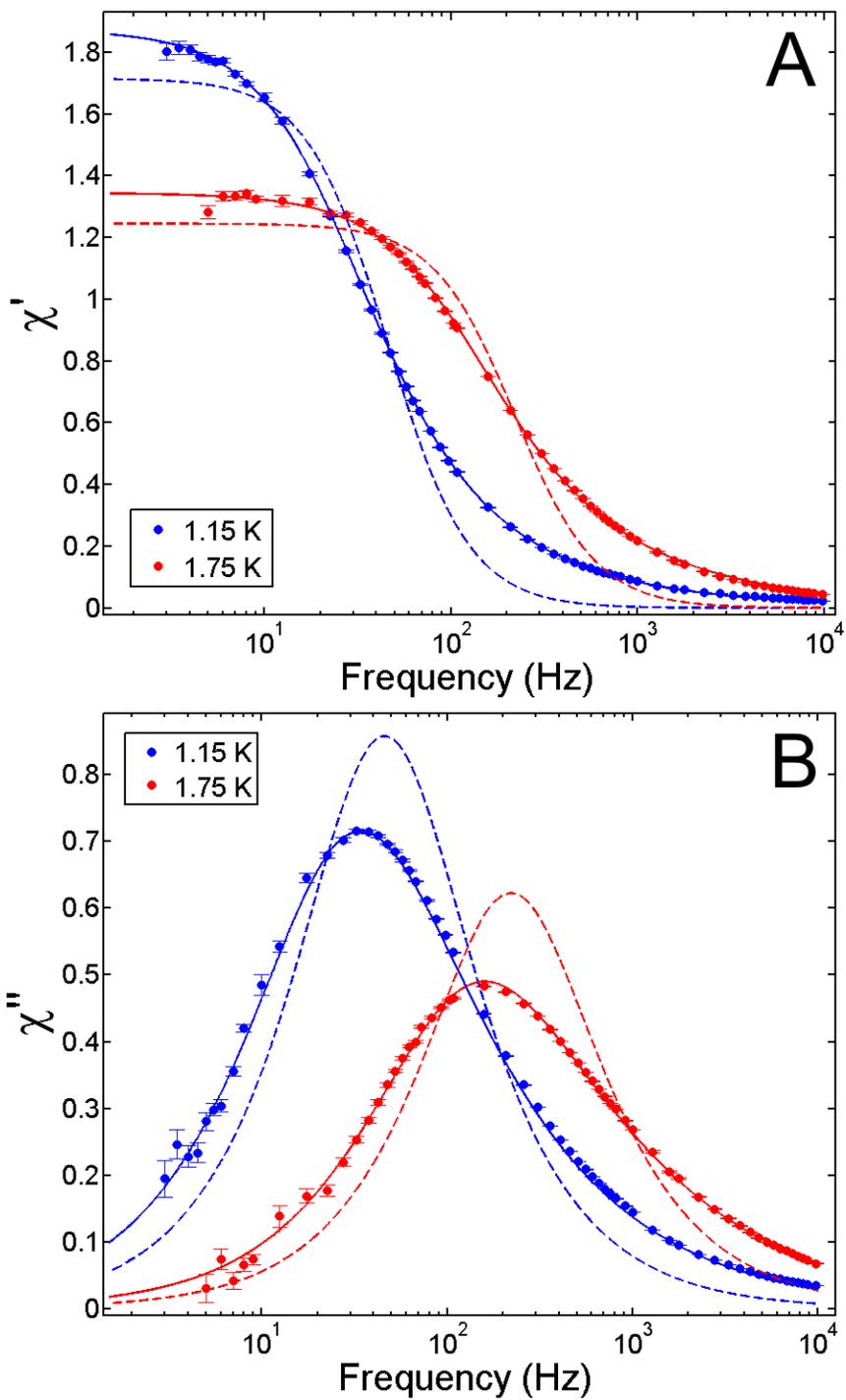

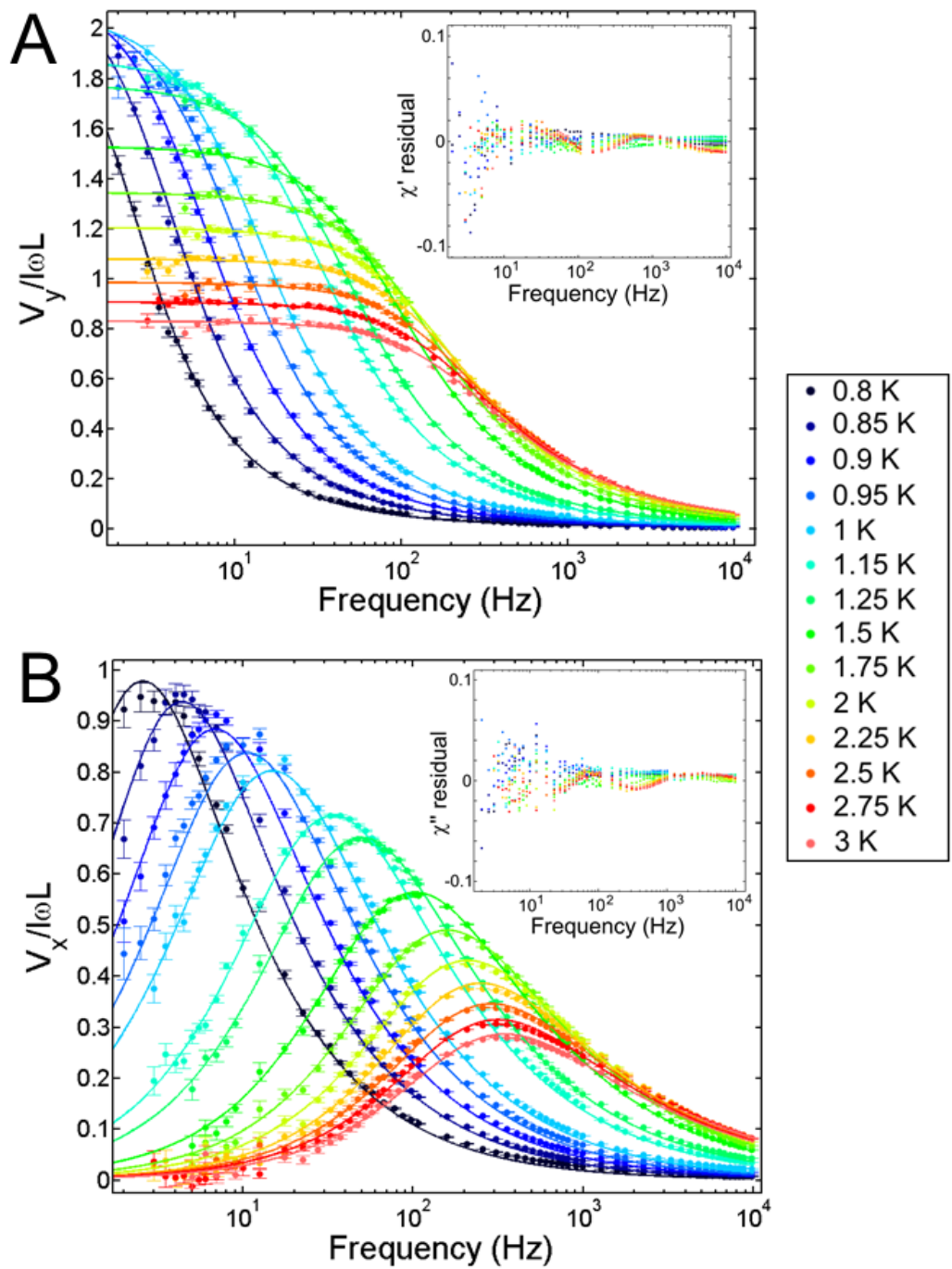

**FIGURE S6**

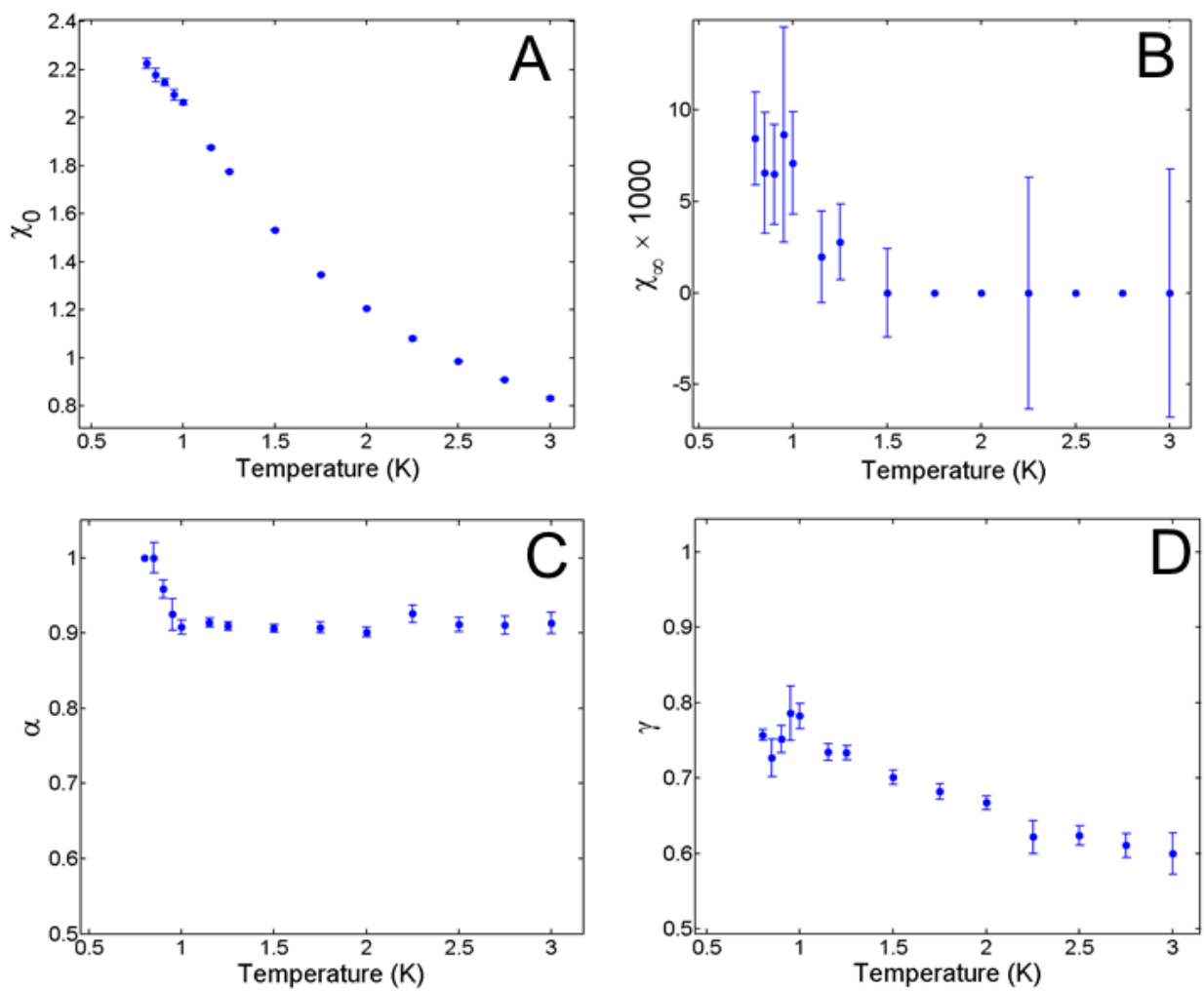

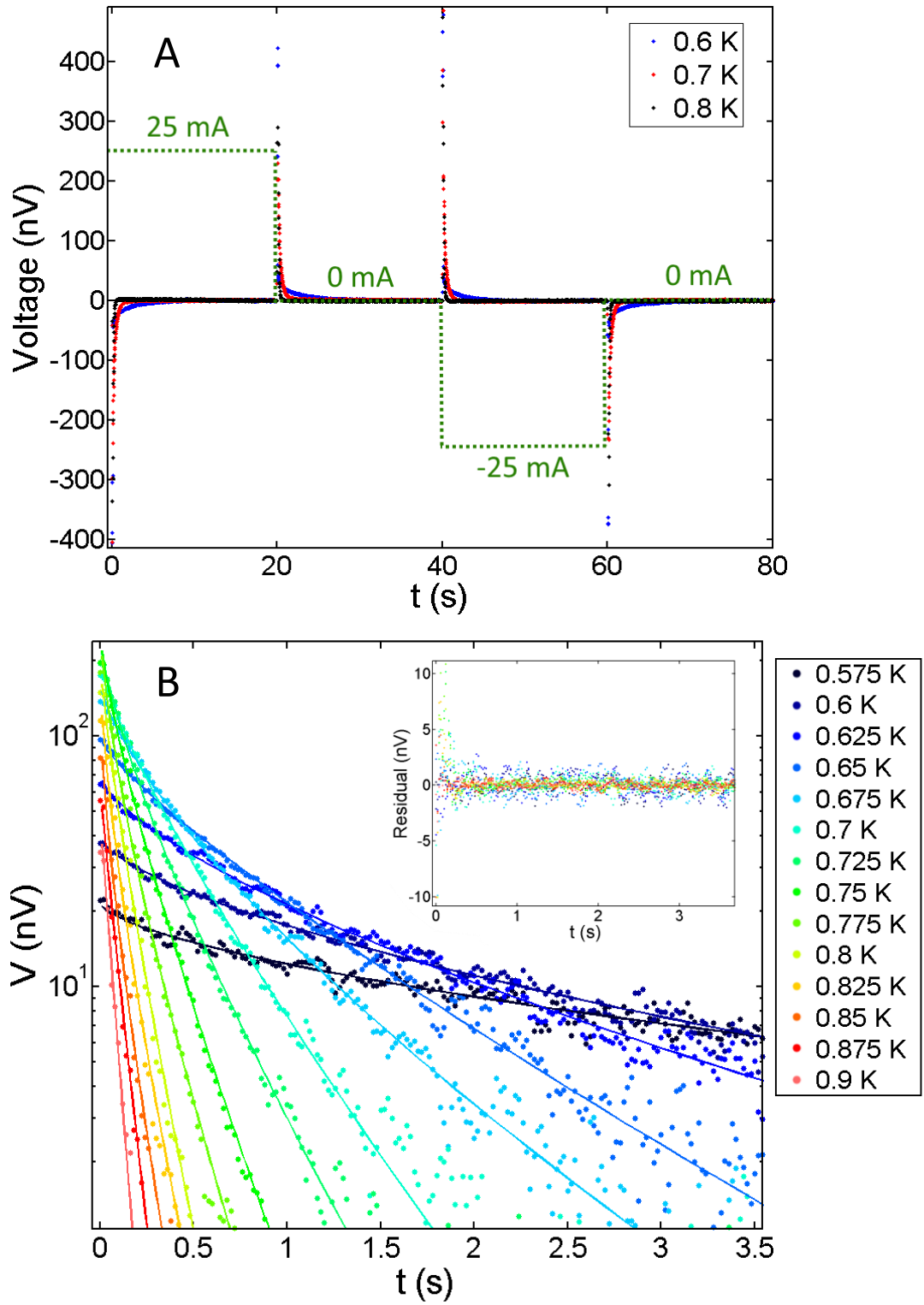

**FIGURE S8**

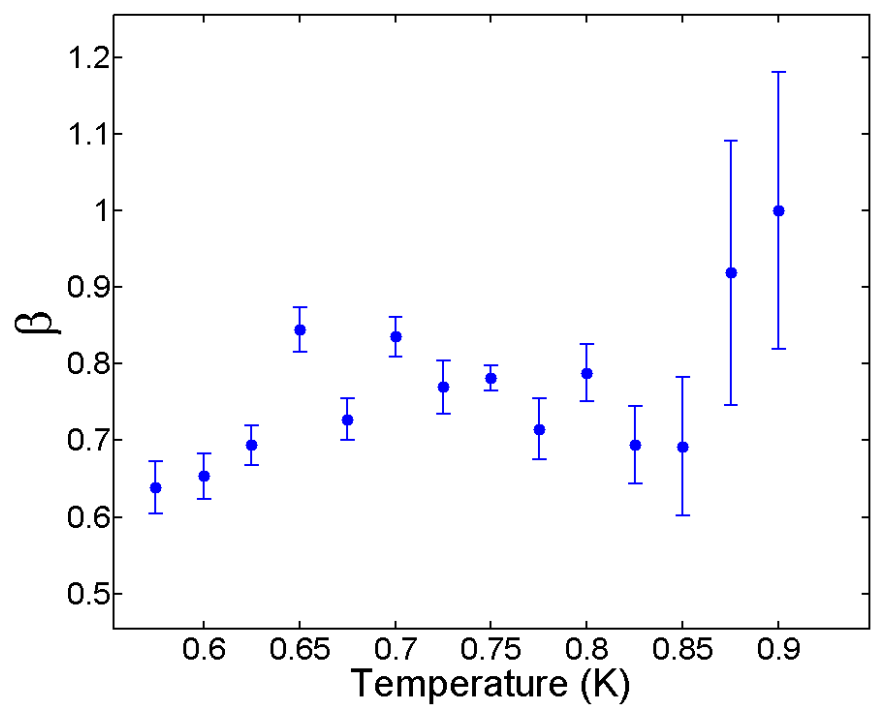

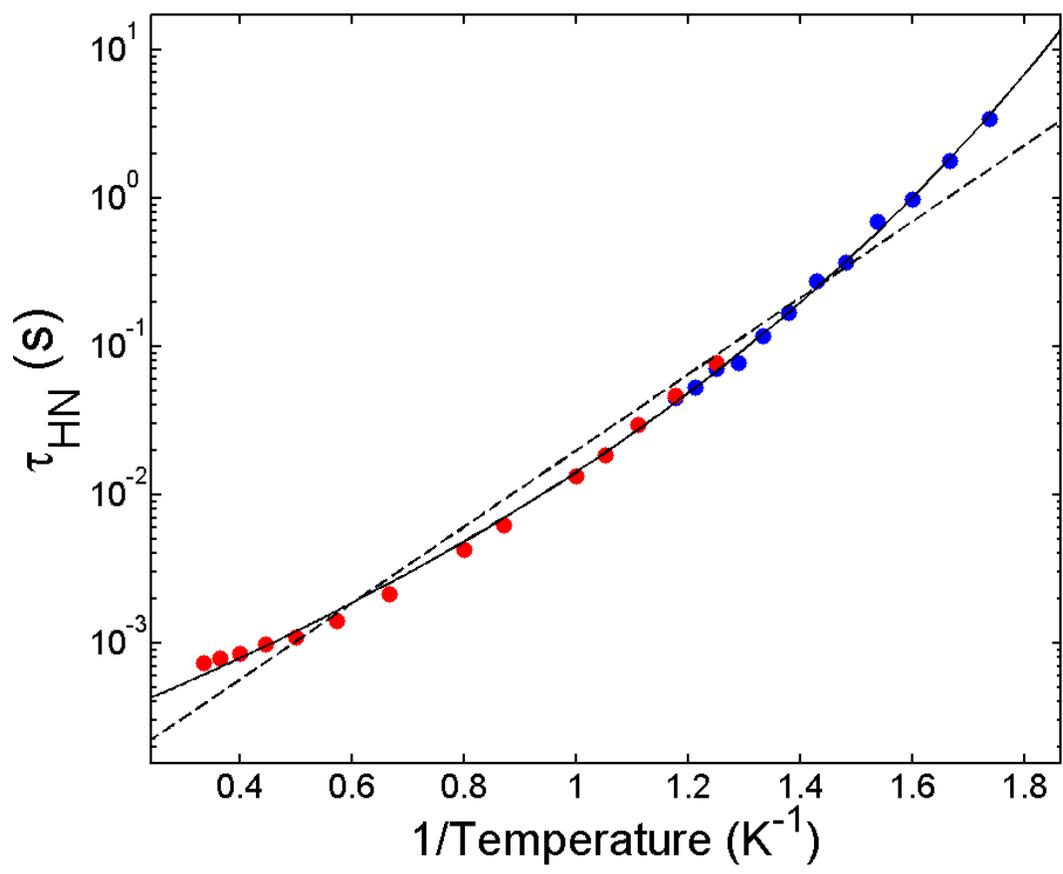

**REFERENCES**

[1] Snyder J *et al.* (2004) Low-temperature spin freezing in the Dy2Ti2O7 spin ice. *Phys. Rev. B* 69:064414.

[2] Yaraskavitch L *et al.* (2012) Spin dynamics in the frozen state of the dipolar spin ice Dy2Ti2O7. *Phys. Rev. B* 85:020410.

[3] Ediger M, Angell C, Nagel S (1996) Supercooled Liquids and Glasses. *J. Phys. Chem.* 100:13200-13212.

[4] Tarjus G, Kivelson S, Nussinov Z, Viot P (2005) The frustration-based approach of supercooled liquids and the glass transition: a review and critical assessment. *J. Phys: Condens. Matt.* 17:R1143-R1182.

[5] Cavagna A (2009) Supercooled liquids for pedestrians. *Physics Reports* 476:51-124.